\documentclass[12pt]{iopart}

\usepackage{graphicx} % Include figure files
\usepackage{color}
\usepackage{tabularx}
\usepackage{doi}

\def\iotabar{\lower3pt\hbox{$\mathchar'26$}\mkern-8mu\iota}
    
\begin{document}
\bibliographystyle{unsrturl}

\title{The impact of rational surfaces on radial heat transport in TJ-II}
\author{B.Ph.~van Milligen$^1$, J.H.~Nicolau$^2$, L. Garc\'ia$^2$, B.A.~Carreras$^2$, C.~Hidalgo$^1$ and the TJ-II Team}
\vspace{3em}
\address{$^1$ CIEMAT - Laboratorio Nacional de Fusi{\'o}n, Avda.~Complutense 40, 28040 Madrid, Spain}
\address{$^2$ Universidad Carlos III, 28911 Legan\'es, Madrid, Spain}
%\address{$^3$ BACV Solutions, 110 Mohawk Road, Oak Ridge, Tennessee 37830, USA}
%\date{\today}

\begin{abstract}
In this work, we study the outward propagation of temperature perturbations.
For this purpose, we apply an advanced analysis technique, the Transfer Entropy, to ECE measurements performed in ECR heated discharges at the low-shear stellarator TJ-II.
We observe that the propagation of these perturbations is not smooth, but is slowed down at specific radial positions, near `trapping zones' characterized by long time lags with respect to the perturbation origin.
We also detect instances of rapid or instantaneous (non-local) propagation, in which perturbations appear to `jump over' specific radial regions.

The analysis of perturbations introduced in a resistive Magneto-Hydrodynamic model of the plasma leads to similar results. 
The radial regions corresponding to slow radial transport are identified with maxima of the flow shear associated with rational surfaces (mini-transport barriers).
The non-local interactions are ascribed to MHD mode coupling effects. 
\end{abstract}


\maketitle

%===========================
\section{Introduction}\label{introduction}

In the quest for energy production based on fusion reactions produced in magnetic confinement devices, so-called Internal Transport Barriers (ITBs) may play an important role.
ITBs are radially localized regions of improved (heat and/or particle) confinement, which may therefore contribute to an overall improvement of confinement and consequently, improved fusion performance~\cite{Wolf:2003,Connor:2004,Garofalo:2015}.

Many factors have an impact on the formation of ITBs, including the local value of the rotational transform ($\iotabar = \iota/2\pi$) or safety factor ($q=2\pi/\iota$), the local magnetic shear, and associated factors such as the current distribution.
ITBs involve a local suppression of turbulence or the radial turbulence correlation length~\cite{Nazikian:2005} and the corresponding turbulent transport, most likely associated with the formation of {local $E_r \times B$ shear layers}~\cite{Diamond:2005}, and hence turbulence modeling will be needed to properly understand this complex phenomenon.
In general, the formation of a velocity shear layer is a function of driving terms such as local (pressure) gradients and damping terms such as viscosity, which may be reduced locally.
However, flow shear layers can also be driven by the turbulence associated with rational $q$ (or $\iotabar$) surfaces~\cite{Diamond:1994}.
%In the case of ITBs, Magneto-HydroDynamic (MHD) modes associated with rational $q$ (or $\iotabar$) surfaces may be involved in the formation of the velocity shear layer~\cite{Thyagaraja:2000}.

Previous experimental studies have shown how rational surfaces and MHD activity are sometimes associated with ITBs for heat transport, in tokamaks~\cite{Koide:1994,Lopes:1997,Mantica:1999,Schilham:2001,Joffrin:2003,Austin:2006}, stellarators~\cite{Hidalgo:2001b,Fujisawa:2002,Ida:2004,Castejon:2004,Estrada:2005,LopezBruna:2008,Itoh:2007}, and RFPs~\cite{Lorenzini:2012}.
The large amount of published work (of which the cited references are only a sample) does not mean, however, that the phenomenon is fully understood or that ITBs can be controlled at will, so further studies are convenient. 
%Recently, it was shown that magnetic islands may be associated with reduced thermal diffusivities~\cite{Ida:2012}.

In this work, we study the relation between rational surfaces and heat transport in the low shear stellarator TJ-II from a novel perspective.
We make use of the excellent external control of the magnetic configuration that this machine affords, particularly in low-$\beta$ Electron Cyclotron Resonance heated plasmas, to place specific rational surfaces at specified radial locations. 
This external control also allows us to improve the statistics of experimental results by averaging over similar discharges.
We then analyze the radial propagation of temperature perturbations using an advanced analysis technique (the Transfer Entropy) that improves upon traditional techniques such as conditional averaging or linear correlation. 
The technique is directional and capable of distinguishing outward from inward propagating perturbations, thus affording greater clarity of results.
Together, these conditions and methods offer an unprecedented view of the impact of rational surfaces on heat transport.  

We will use a resistive Magneto-HydroDynamic model to interpret and support some of the reported experimental observations. 

This paper is structured as follows. 
In Section \ref{techniques}, the experimental setup is discussed.
In Section \ref{experiments}, we present the experimental results.
In Section \ref{modeling}, we present the modeling results.
In Section \ref{discussion}, we discuss the results, and draw some conclusions.

%===========================
\clearpage
\section{Experimental set-up and techniques used}\label{techniques}

In this work, we study discharges heated by Electron Cyclotron Resonant Heating (ECRH).
In these discharges, the plasma has a relatively low line average electron density of $\overline n_e \simeq 0.5 \cdot 10^{19}$ m$^{-3}$, mostly below the critical density of the electron to ion root confinement transition at TJ-II~\cite{Hidalgo:2006b,Velasco:2012}.
%This also means that Neoclassical effects are not expected to play a significant role in the explanation of the observed barriers.

\subsection{Magnetic configurations}

TJ-II is a stellarator of the heliac type with a major radius of $R_0 = 1.5$ m, a minor radius of $a \simeq 0.2$ m, and four field periods~\cite{Alejaldre:1999}.
The TJ-II vacuum magnetic geometry is completely determined by the currents flowing in the external coil sets. 
The magnetic field is normalized to 0.95 T on the magnetic axis at the ECRH injection point in order to guarantee central absorption of the ECR heating power~\cite{Milligen:2011c}.
In ECRH heated discharges at TJ-II, the normalized pressure $\left < \beta \right >$ remains rather low and currents flowing inside the plasma are generally quite small (unless explicitly driven), so that the actual magnetic configuration is typically rather close to the vacuum magnetic configuration~\cite{Milligen:2011b}.

\begin{figure}\centering
  \includegraphics[trim=0 0 0 0,clip=,width=16cm]{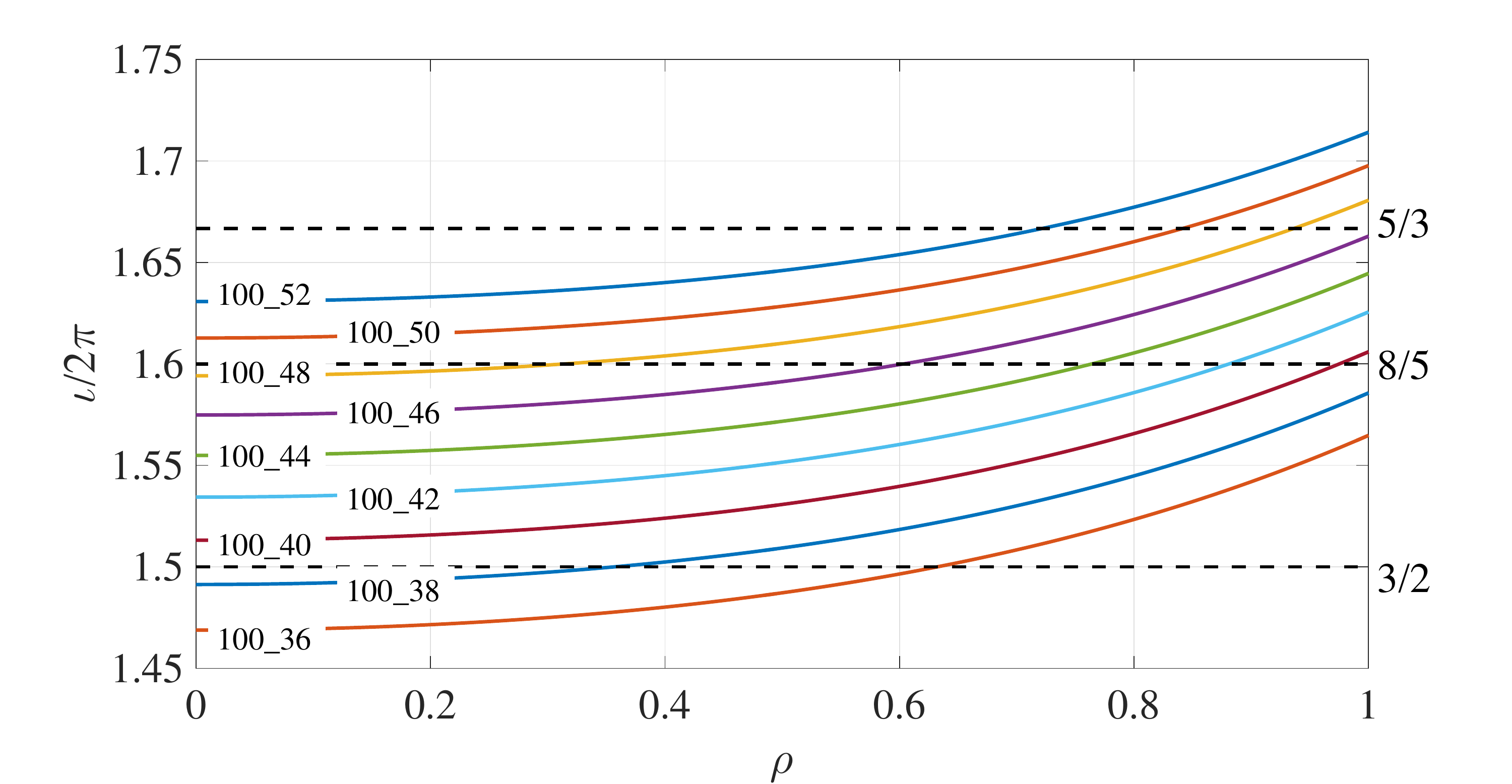}
\caption{\label{iota}Profiles of the rotational transform, $\iotabar = \iota/2\pi$ as a function of normalized radius, $\rho = r/a$. Major rational values are indicated by horizontal dashed lines. The line labels are identifiers of the magnetic configuration corresponding to each $\iotabar$ profile.}
\end{figure}

Thus, the external coil currents provide exquisite control of the magnetic configuration, making it possible to average experimental results over similar discharges obtained in the same magnetic configuration.
In this paper, we use a set of configurations that is rather similar regarding field strength and plasma shape, but with different profiles of the rotational transform, $\iotabar = \iota/2\pi$, as shown in Fig.~\ref{iota}.
This variation in rotational transform profiles means that specific rational surfaces are located at different radial positions in different configurations, permitting the study of the impact of these rational surfaces on plasma properties.

\subsection{ECRH and ECE measurements}\label{ecrh}

The ECRH system consists of two gyrotrons with a frequency of 53.2 GHz, allowing the injection of up to $2 \times 300$ kW of heating power~\cite{Fernandez:2009}.
In the ECRH experiments, both ECRH systems were typically launching a power of 250 kW each into the core of the plasma; the half width of the deposition profile being $w_{\rm ECRH} \simeq 3$ cm~\cite{Eguilior:2003}. 

TJ-II disposes of a 12 channel Electron Cyclotron Emission (ECE) detection system to measure the local electron temperature $T_e$ at up to 12 different radial positions along the midplane on the high magnetic field side of the plasma (at a toroidal angle of $\phi = 315^\circ$), covering a significant part of the plasma minor radius (about 70\% of the minor radius), with a radial resolution of about 1 cm~\cite{delaLuna:2001}. 
The ECE channels are tuned to the second harmonic of the electron cyclotron frequency at various positions inside the plasma.
For each magnetic configuration, the radial position corresponding to a given ECE channel, $\rho(i)$, is determined as follows. 
The magnetic field distribution $B(\vec R)$ is calculated from the currents flowing through the external field coils.
By equating the second harmonic of the electron cyclotron frequency, $\omega_{ce} = eB/m$, to the measurement frequency of the channel, $2\pi f(i)$, one then obtains the space coordinates $\vec R(i)$ corresponding to the observation point of the channel along the line of sight of the diagnostic.
From these coordinates, the corresponding value of the poloidal magnetic flux is calculated, using the known poloidal magnetic flux distribution in vacuum, $\psi(\vec R(i))$, which is directly related to the radial positions as $\rho \propto \sqrt{\psi}$ by definition.
By convention, positive $\rho$ values correspond to the low field side of the plasma and negative $\rho$ values to the high field side.

\subsection{Temperature fluctuations and propagation of perturbations}

\begin{figure}\centering
  \includegraphics[trim=0 0 0 0,clip=,width=16cm]{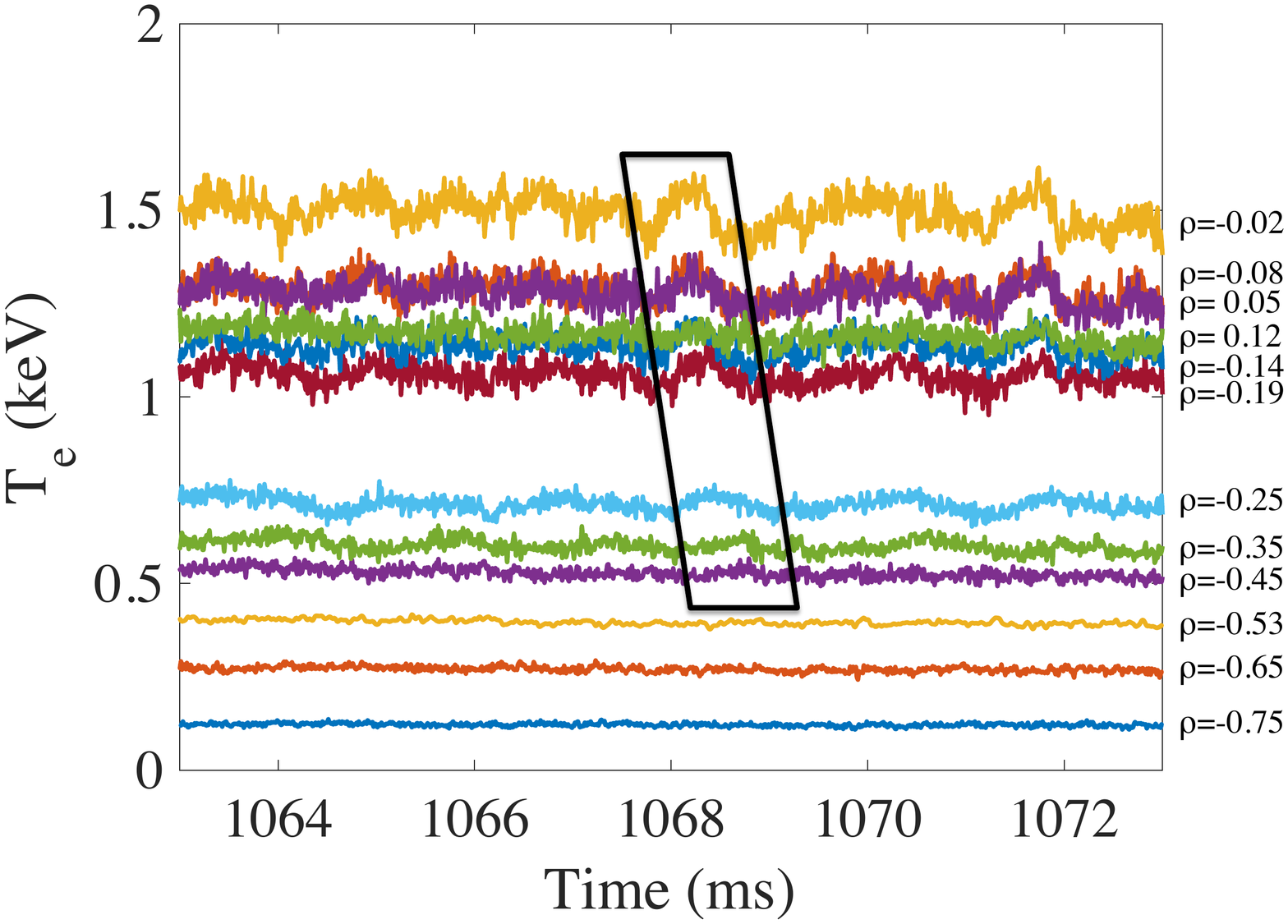}
\caption{\label{rawdata}Example of ECE data. Labels indicate the radial position ($\rho = r/a$) of the measurement. The parallelepiped highlights the propagation of a single spontaneously generated heat pulse from the core to the edge.}
\end{figure}

The ECRH power deposited in the core of the plasma causes spontaneous temperature fluctuations~{\cite{Happel:2015,Hidalgo:2016}}, possibly related to the presence of rational surfaces in the core region of the plasma where the heat is deposited~\cite{Estrada:2002} and the concomitant generation of fast electrons~\cite{Garcia:2016}. These temperature fluctuations then lead to the outward propagation of small corresponding cascades or perturbations (cf. Fig.~\ref{rawdata}), similar to what has been reported in Ref.~\cite{Politzer:2000}.
In this work, we will exploit this phenomenon to analyze heat transport outside the core power deposition region.

\subsection{Transfer Entropy}\label{transferentropy}

To analyze these small amplitude, randomly occurring propagating temperature fluctuations, traditional heat pulse analysis techniques (correlation, conditional averaging~\cite{Teliban:2007}) do not offer sufficient clarity:
the correlation tends to smear out the information over the typical duration of the heat pulses, while the conditional averaging technique tends to offer poor statistics due to the necessity to clearly identify individual (relatively large) events.

For this reason, we turn to a technique from the field of Information Theory~\cite{Schreiber:2000} that was recently applied for the first time in the context of fusion plasmas~\cite{Milligen:2014}: the Transfer Entropy.
This nonlinear technique measures the `information transfer' between two signals, is directional, and uses all the information available in the two signals, regardless of amplitude or sign. 

The Transfer Entropy is a measure of the causal relation or information flow between two time series. 
It is based on the concept of `quantifiable causality' introduced by Wiener~\cite{Wiener:1956} (rephrased slightly): 
{\it For two simultaneously measured signals X and Y, if we can predict X better by using the past information from Y than without it, then we call Y causal to X.}
Accordingly, the 
Transfer Entropy between signals $Y$ and $X$ quantifies the number of bits by which the prediction of a signal $X$ can be improved by using the time history of not only the signal $X$ itself, but also that of signal $Y$.

Consider two processes $X$ and $Y$ yielding discretely sampled time series data $x_i$ and $y_j$.
In this work, we use a simplified version of the Transfer Entropy:
\begin{equation}\label{TE}
T_{Y \to X} = \sum{p(x_{n+1},x_{n-k},y_{n-k}) \log_2 \frac{p(x_{n+1}|x_{n-k},y_{n-k})}{p(x_{n+1}|x_{n-k})}} .
\end{equation}
Here, $p(a|b)$ is the probability distribution of $a$ conditional on $b$, $p(a|b) = p(a,b)/p(b)$.
The probability distributions $p(a,b,c,\dots)$ are constructed using $m$ bins for each argument, i.e., the object $p(a,b,c,\dots)$ has $m^d$ bins, where $d$ is the dimension (number of arguments) of $p$. 
The sum in Eq.~\ref{TE} runs over the corresponding discrete bins.
The number $k$ can be converted to a `time lag' by multiplying it by the sampling rate.
The construction of the probability distributions is done using `course graining', i.e., a low number of bins (here, $m=3$), to obtain statistically significant results. For more information on the technique, please refer to Ref.~\cite{Milligen:2014}. 
The value of the Transfer Entropy $T$, expressed in bits, can be compared with the total bit range, $\log_2 m$, equal to the maximum possible value of $T$, to help decide whether the Transfer Entropy is significant or not.

\begin{figure}\centering
  \includegraphics[trim=0 0 0 0,clip=,width=12cm]{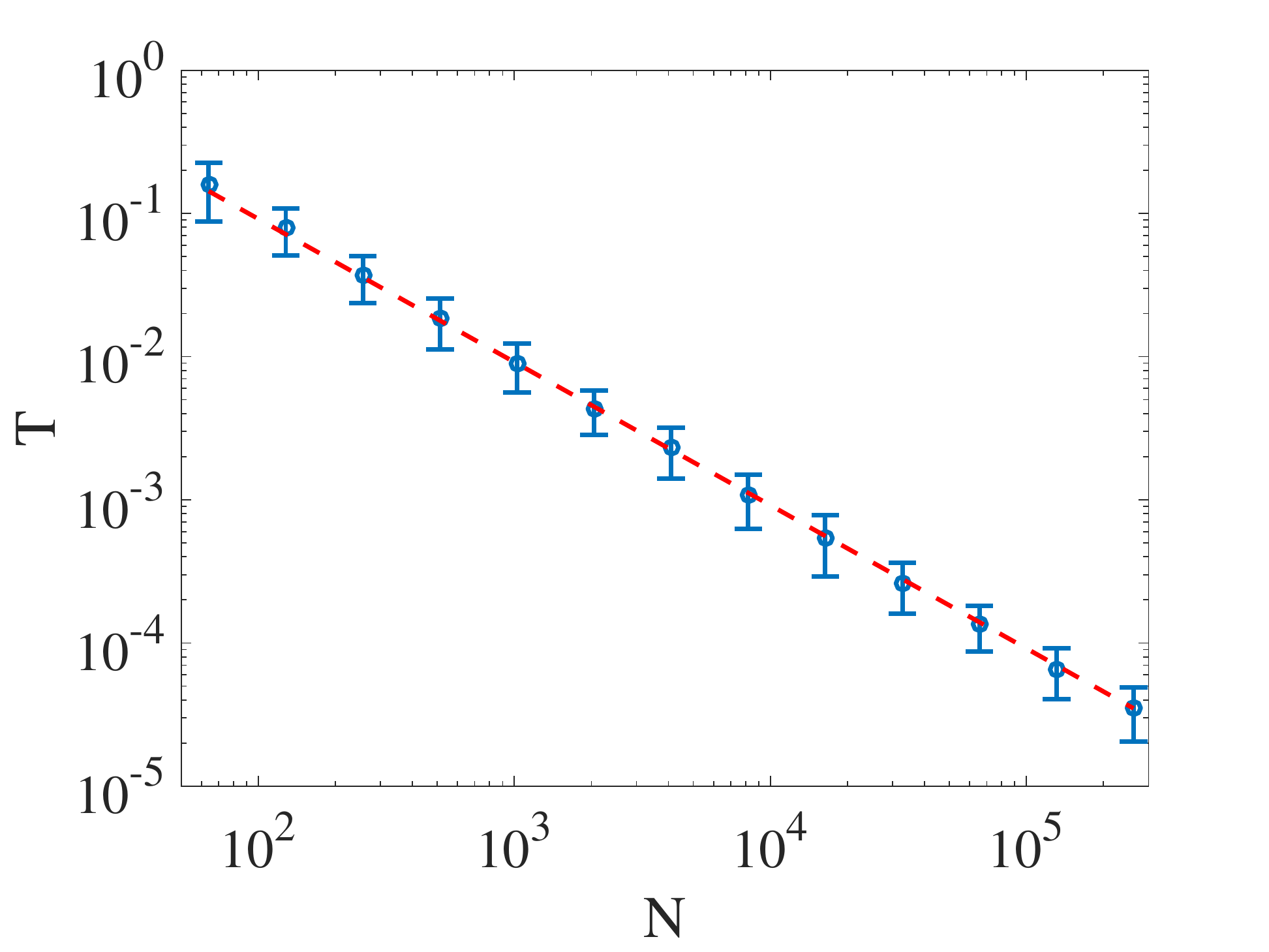}
\caption{\label{TE_random}{Transfer Entropy for two random Gaussian signals, as a function of the number of samples of the signals, $N$ (with $m=3$). 
Each point is calculated as the average over 100 independent realizations, and the error bar indicates the variation of the result.
This average value can be taken as the statistical significance level of the Transfer Entropy. The red dashed line is proportional to $1/N$.}}
\end{figure}

A simple way of estimating the statistical significance of the Transfer Entropy is by calculating $T$ for two random (noise) signals.
Fig.~\ref{TE_random} shows the Transfer Entropy calculated for two such random signals, with a Gaussian distribution, each with a number of samples equal to $N$.
It can be seen that the value of the Transfer Entropy (averaged over 100 equivalent realizations) drops proportionally to $1/N$~ \cite{Milligen:2016c}.
Here, we will be analyzing signals with a typical duration between 50 and 150 ms, corresponding to $5\cdot 10^3 \le N \le 1.5 \cdot 10^4$ points, so that the statistical significance level of $T$ is of the order of $10^{-3}$ or less.

Regarding the interpretation of the Transfer Entropy, we note that it is a non-linear quantifier of information transfer and helps clarifying which fluctuating variables influence which others - without specifying the nature of this influence. 
In this sense, it is fundamentally different from the cross correlation, which is maximal for two identical signals ($X=Y$), whereas the Transfer Entropy is exactly zero for two identical signals (as no information is gained by using the second, identical signal to help predicting the behavior of the first).
Furthermore, the Transfer Entropy is {\it directional} and therefore capable of distinguishing perturbations propagating from $Y$ to $X$ or vice versa.
Another competing technique is the `conditional average', in which a (threshold) condition is applied to one signal to provide a trigger time for averaging the second one. This simple and easily understood technique is very useful, provided the condition is clear, sharp, and unambiguous~\cite{Teliban:2007}. By comparison, the Transfer Entropy is more robust and independent from arbitrary external conditions such as the choice of threshold.

\subsection{Averaging procedure}\label{averaging}

In Section \ref{experiments}, we will analyze experimental data obtained from TJ-II.
We are mainly interested in the effect of rational surfaces on heat transport.
As noted, the location of the rational surfaces is controlled precisely.
Individual plasmas experience variations due to changes in ECRH heating power, gas puff, wall conditions, impurity content and radiation levels, turbulence, etc.
We use ECE measurements to track outward propagating temperature perturbations.
Slow data drifts are removed by applying a high-pass filter with 100 Hz cutoff prior to analysis.
The analysis technique we apply (the Transfer Entropy between a central reference ECE channel and other ECE channels) allows us to track the outward propagation of perturbations in terms of the ability to predict temperature fluctuations at a given radius, given knowledge about temperature fluctuations in the core region.
This somewhat abstract criterion is fully insensitive to confounding details like data amplitude or sign (and therefore, data calibration), as well as the shape or amplitude of propagating perturbations: only the relative time delay matters.
The analysis offers another advantage: since it is {\it directional}, one may select outward propagating events, ignoring inward propagating events (which may occur simultaneously~\cite{Milligen:2002}).

In this framework, it will be fruitful to average the calculated Transfer Entropy results over similar discharges having the same magnetic configuration.
Shot to shot variations in the Transfer Entropy, not systematically related to the magnetic configuration, will be suppressed by the averaging procedure.
On the other hand, effects associated with the magnetic configuration will be emphasized. 

%===========================
\clearpage
\section{Experiments}\label{experiments}

\begin{figure}\centering
  \includegraphics[trim=0 0 0 0,clip=,width=16cm]{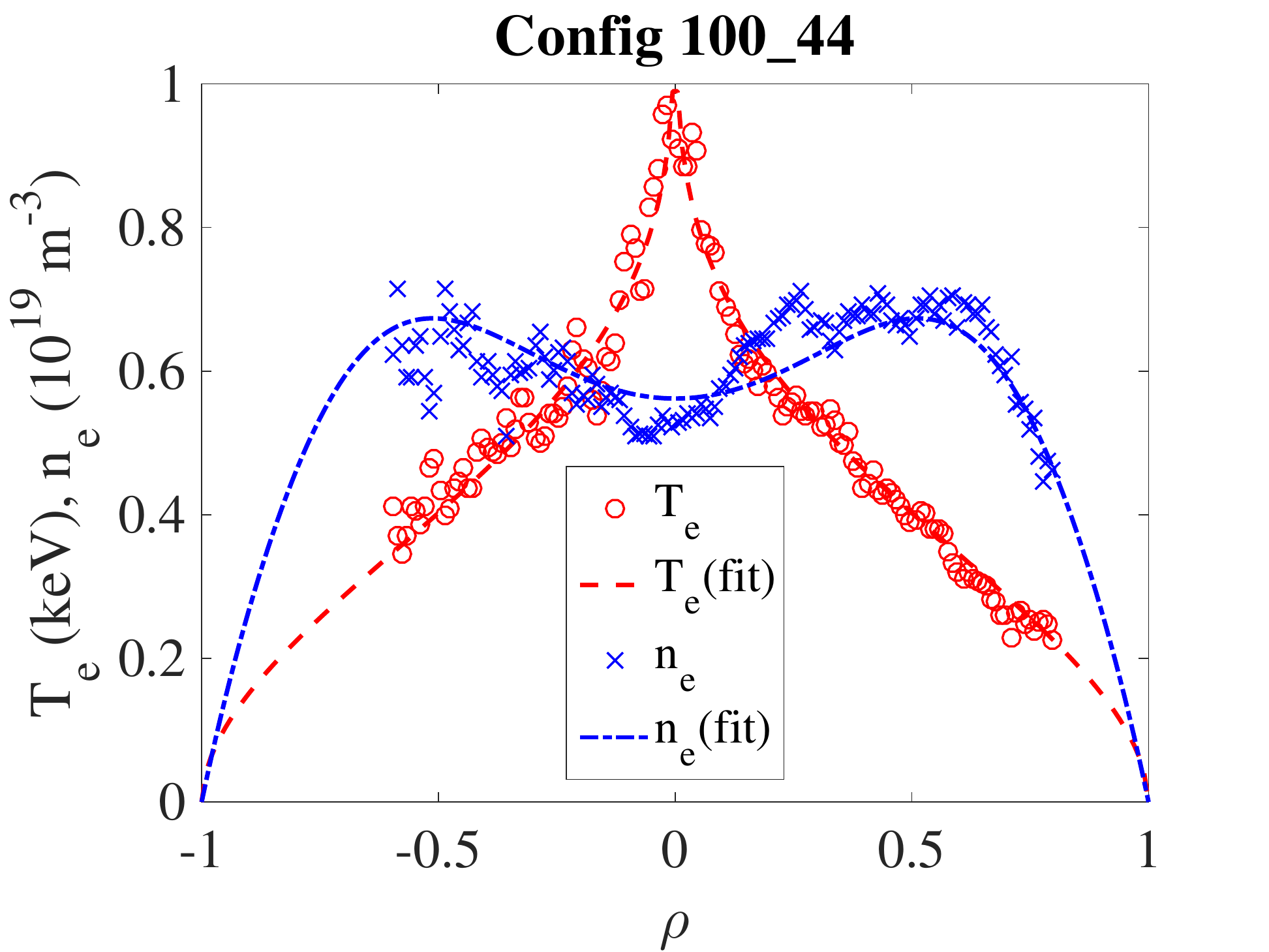}
\caption{\label{TSprof_100_44}{Typical Thomson Scattering profiles for ECRH discharges in TJ-II (data have been averaged over 9 discharges).}}
\end{figure}

Fig.~\ref{TSprof_100_44} shows typical Thomson Scattering~\cite{Barth:1999} profiles for ECRH discharges in TJ-II.
The electron density profile, $n_e$,  is broad and slightly hollow, while the electron temperature profile, $T_e$, is peaked due to ECRH power deposition in the plasma center~{\cite{Ascasibar:2005,Milligen:2003}}.  
These profiles {change very little for the discharges} studied here. Configurations are fixed in each discharge studied. 

\subsection{Spontaneously generated propagating temperature perturbations}

In this section, we analyze experiments with continuous central ECR heating. 
As noted, spontaneous electron temperature fluctuations occur in the plasma core region, leading to outward propagating temperature perturbations.
Since this type of experiments is common at TJ-II, the database of available discharges is very large and includes many magnetic configurations.
We have analyzed measurements of $T_e(\rho,t)$ obtained from the ECE diagnostic during the steady state phase of the discharges.
The length of the steady state phase varied between about 50 and 150 ms. 
We calculated the Transfer Entropy from the most centrally located channel (the one having the smallest value of $|\rho|$, taken to be the `reference' channel) to the other channels, for a range of time lag values $\tau$. 
This allows visualizing the propagation of spontaneously arising temperature fluctuations from the center towards the edge of the plasma, regardless of the amplitude or shape of these propagating perturbations; only the relative time delay of the `information transfer' is relevant.

\begin{figure}\centering
  \includegraphics[trim=50 175 50 25,clip=,width=16cm]{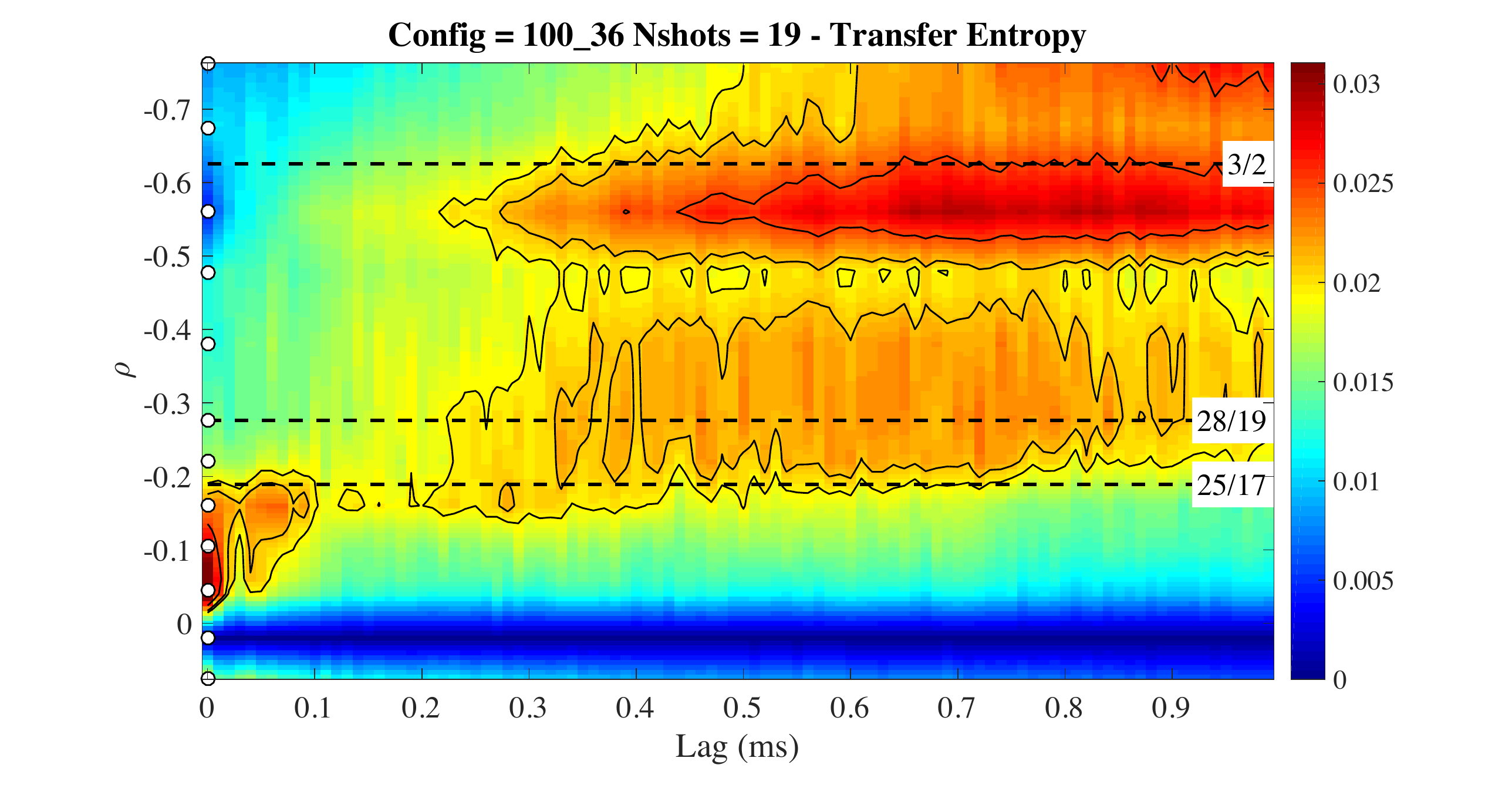}
\caption{\label{mean_TE_100_36}{Transfer Entropy $T(\rho,\tau)$ for configuration 100\_36 (see Fig.~\ref{iota}).
Temperature perturbations originating from the core appear to be held up near the $25/17$ rational surface, immediately outside the core region. After a lag of about 0.2 ms, the perturbations `break through' this barrier, reaching $|\rho| \simeq 0.3$ and almost simultaneously $|\rho| \simeq 0.55$, so that the perturbations seem to `jump over' the intermediate zone.
}}
\end{figure}

\begin{figure}\centering
  \includegraphics[trim=50 175 50 25,clip=,width=16cm]{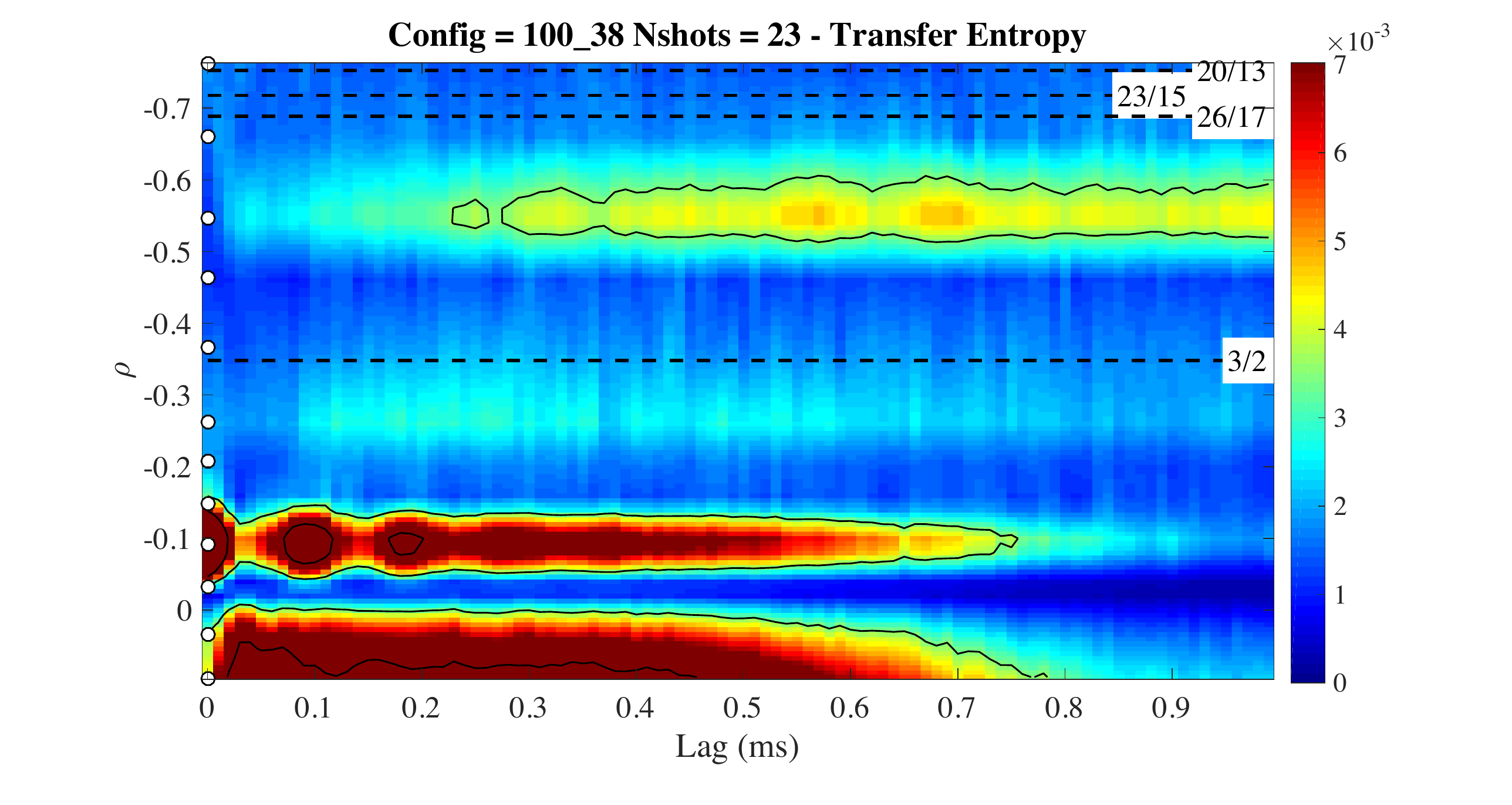}
\caption{\label{mean_TE_100_38}{Transfer Entropy $T(\rho,\tau)$ for configuration 100\_38 (see Fig.~\ref{iota}).
Temperature perturbations almost do not propagate in a rather wide region, $0.2 < |\rho| < 0.5$, around the $3/2$ rational surface. The core perturbations do have a delayed effect at $|\rho| \simeq 0.55$, so that the perturbations seem to `jump over' the influence region of the $3/2$ rational surface.
}}
\end{figure}

\begin{figure}\centering
  \includegraphics[trim=50 175 50 25,clip=,width=16cm]{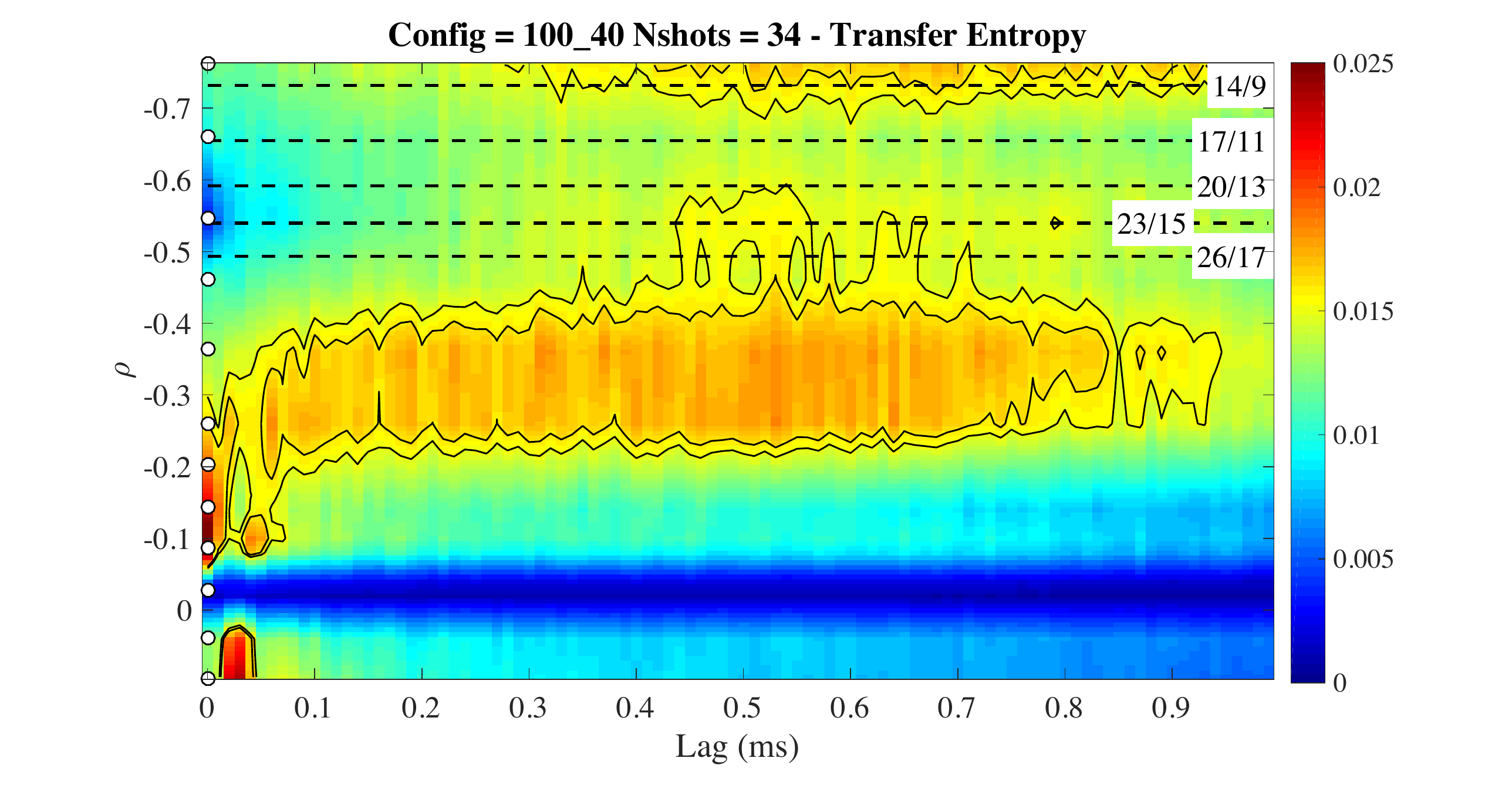}
\caption{\label{mean_TE_100_40}{Transfer Entropy $T(\rho,\tau)$ for configuration 100\_40 (see Fig.~\ref{iota}).
Temperature perturbations propagate rapidly from the core region to around $|\rho| \simeq 0.3$ (in about 0.1 ms), but then are detained, presumably due to the presence of the set of rational surfaces occurring at $|\rho| > 0.5$.
}}
\end{figure}

\begin{figure}\centering
  \includegraphics[trim=50 175 50 25,clip=,width=16cm]{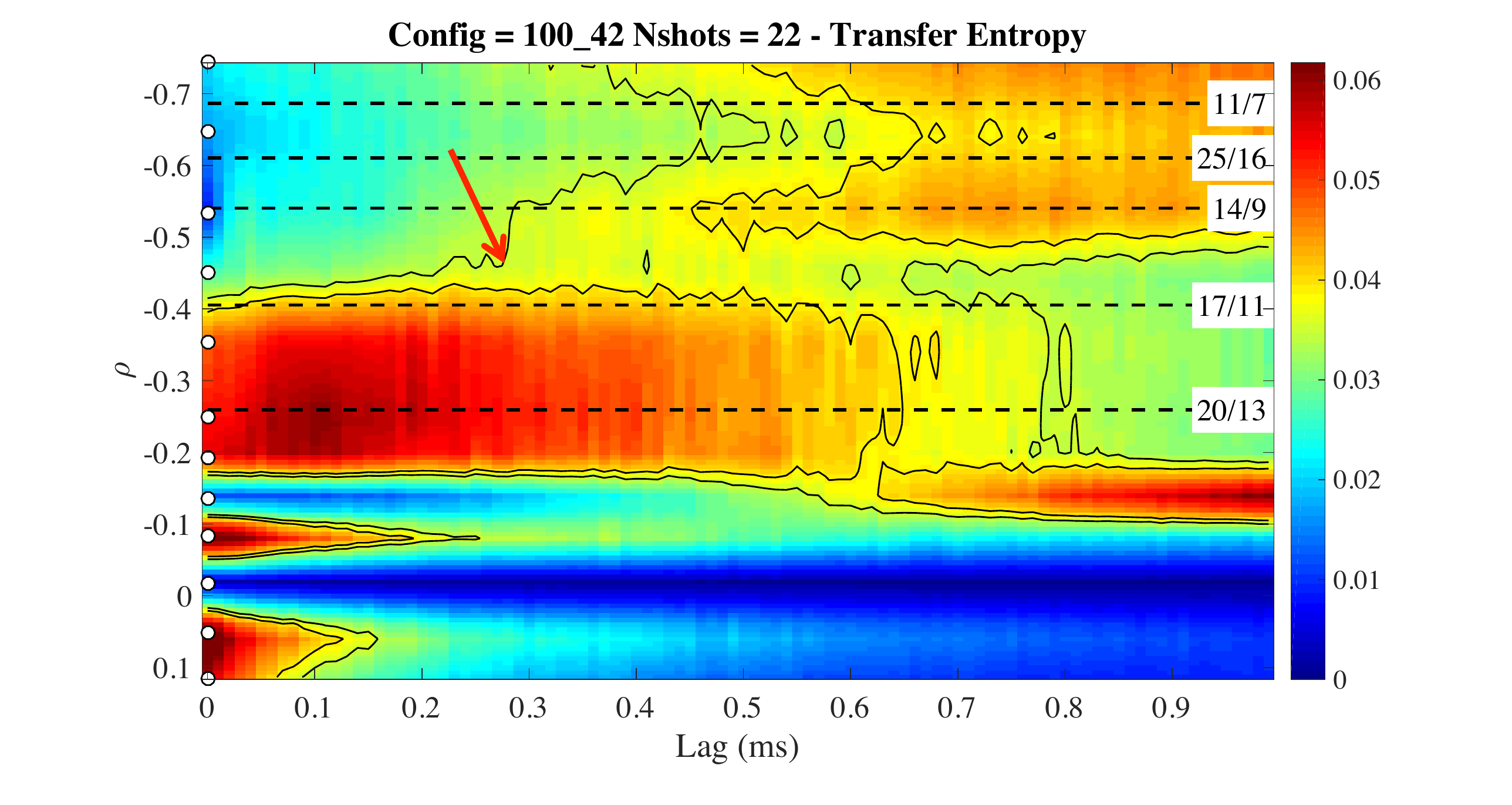}
\caption{\label{mean_TE_100_42}{Transfer Entropy $T(\rho,\tau)$ for configuration 100\_42 (see Fig.~\ref{iota}).
Temperature perturbations propagate rapidly from the core region to the approximate location of the $17/11$ rational surface. There, they are detained although eventually, at lags between 0.3 and 0.5 ms, the perturbations `break through' this barrier and reach $|\rho| \simeq 0.55$.
The arrow indicates a `step' in the propagation front.
}}
\end{figure}

\begin{figure}\centering
  \includegraphics[trim=50 175 50 25,clip=,width=16cm]{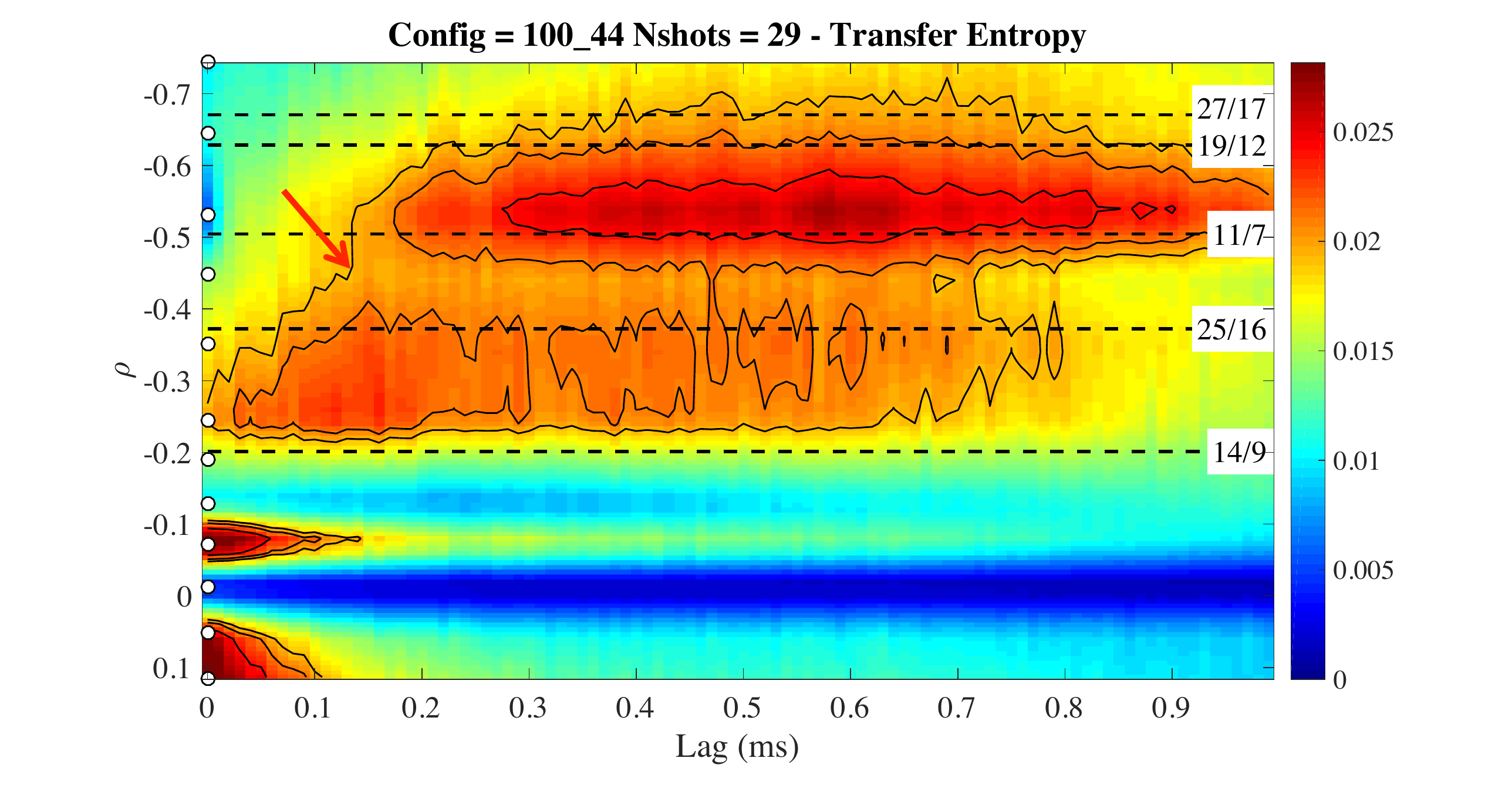}
\caption{\label{mean_TE_100_44}{Transfer Entropy $T(\rho,\tau)$ for configuration 100\_44 (see Fig.~\ref{iota}).
Outward temperature perturbations encounter a barrier near the 25/16 rational surface, although they `break through' rapidly, at lags around 0.2 ms, and reach $|\rho| \simeq 0.55$.
The arrow indicates a `step' in the propagation front.
}}
\end{figure}

\begin{figure}\centering
  \includegraphics[trim=50 175 50 25,clip=,width=16cm]{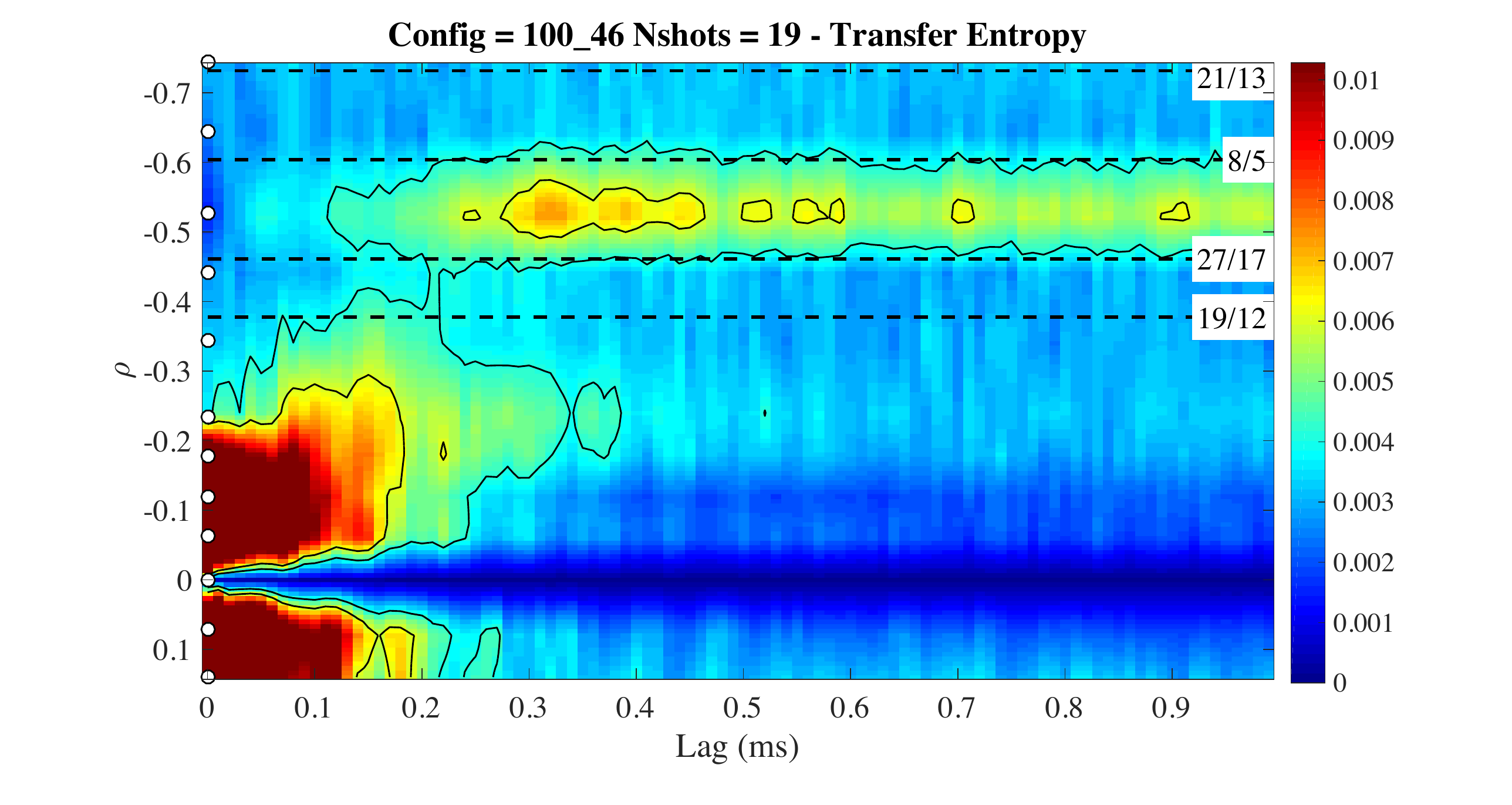}
\caption{\label{mean_TE_100_46}{Transfer Entropy $T(\rho,\tau)$ for configuration 100\_46 (see Fig.~\ref{iota}).
Outward temperature perturbations encounter a barrier near the 19/12 rational surface, although they `break through' rapidly, at lags around 0.2 ms, and reach $|\rho| \simeq 0.55$.
}}
\end{figure}

\begin{figure}\centering
  \includegraphics[trim=50 175 50 25,clip=,width=16cm]{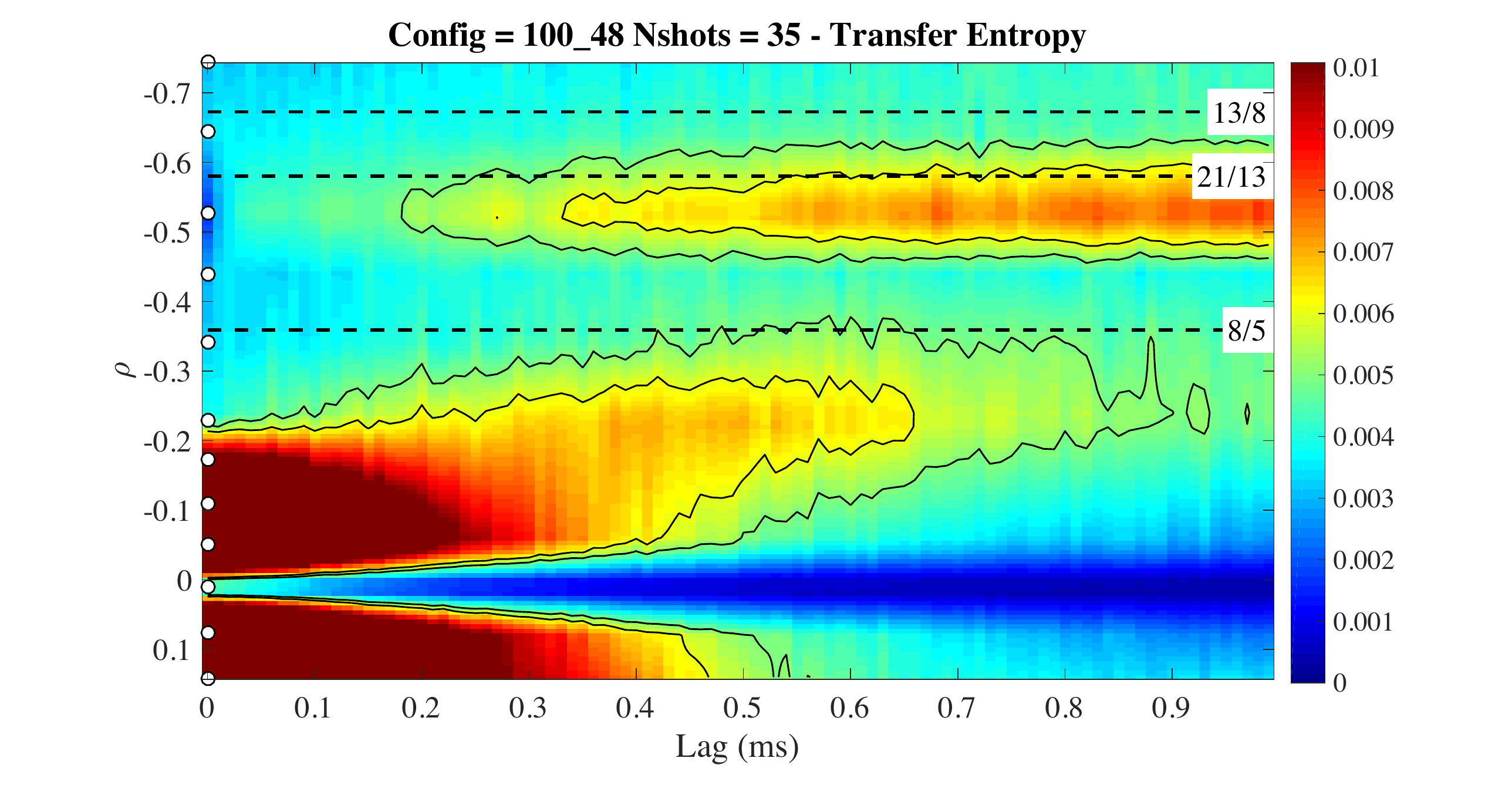}
\caption{\label{mean_TE_100_48}{Transfer Entropy $T(\rho,\tau)$ for configuration 100\_48 (see Fig.~\ref{iota}).
Outward temperature perturbations encounter a strong barrier near the 8/5 rational surface, due to which the outward propagation velocity of the temperature perturbations is quite low. Even so, the core temperature perturbations also have a delayed effect at $|\rho| \simeq 0.55$, so that the perturbations seem to `jump over' the influence region of the $8/5$ rational surface.
}}
\end{figure}

\begin{figure}\centering
  \includegraphics[trim=50 175 50 25,clip=,width=16cm]{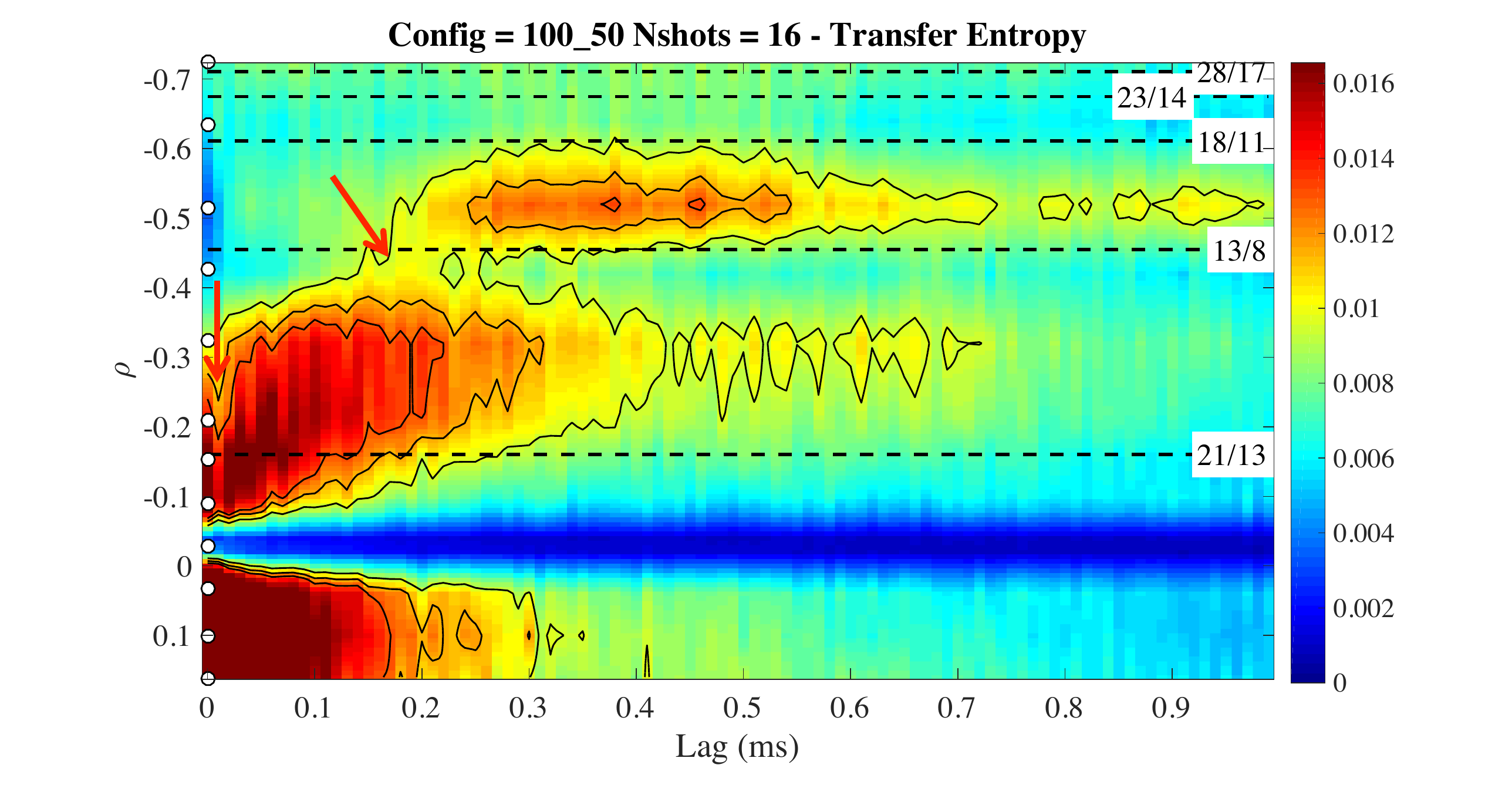}
\caption{\label{mean_TE_100_50}{Transfer Entropy $T(\rho,\tau)$ for configuration 100\_50 (see Fig.~\ref{iota}).
Outward propagating temperature perturbations encounter a barrier near the 13/8 rational surface, although they `break through' rapidly, at lags around 0.2 ms, and reach $|\rho| \simeq 0.55$.
The arrow indicates a `step' in the propagation front.
}}
\end{figure}

\begin{figure}\centering
  \includegraphics[trim=50 175 50 25,clip=,width=16cm]{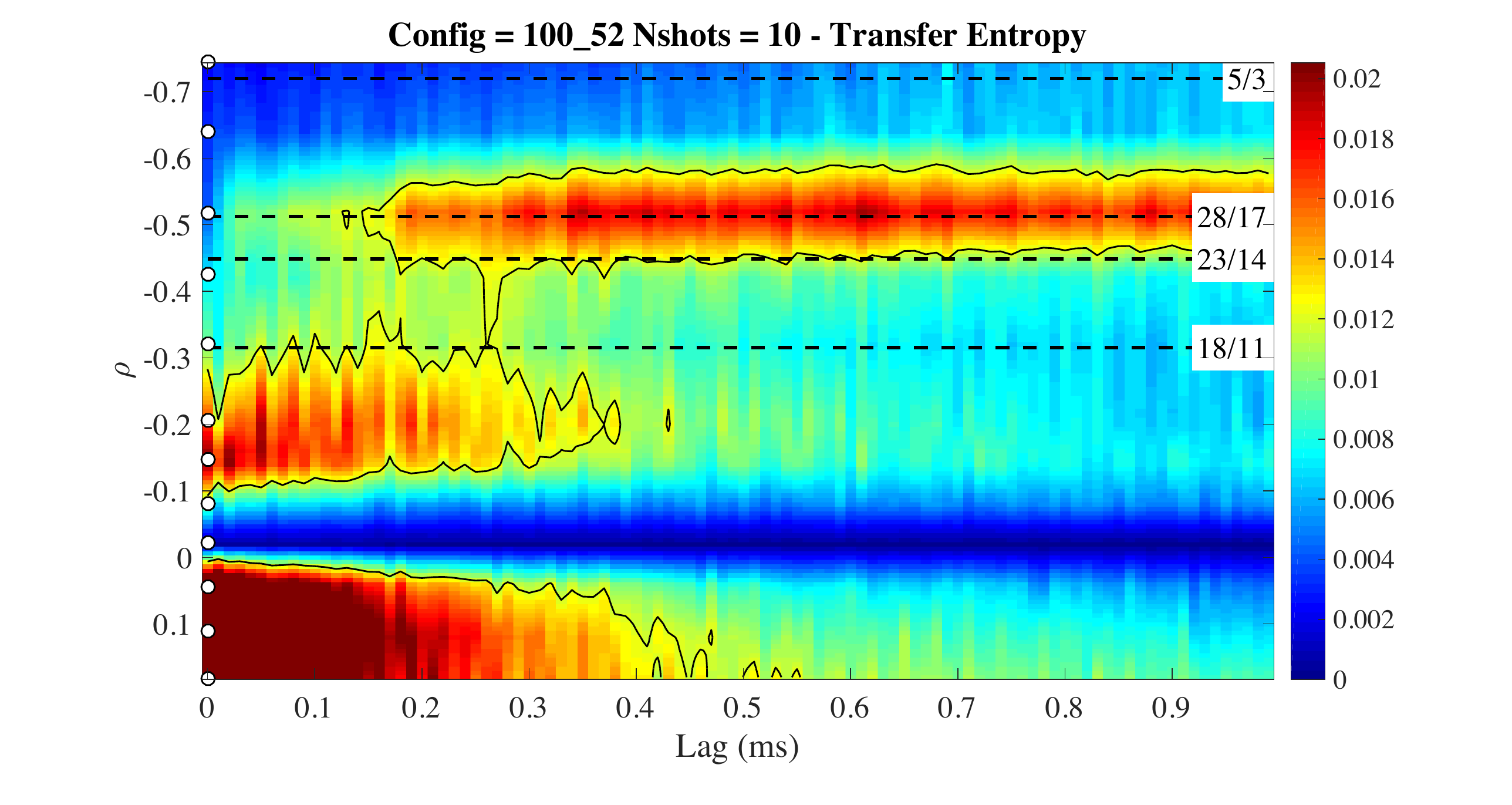}
\caption{\label{mean_TE_100_52}{Transfer Entropy $T(\rho,\tau)$ for configuration 100\_52 (see Fig.~\ref{iota}).
Outward propagating temperature perturbations encounter a relatively strong barrier near the 18/11 rational surface, due to which the outward propagation velocity of the temperature perturbations is quite slow. Even so, the core temperature perturbations also have a delayed effect at $|\rho| \simeq 0.55$, so that the perturbations seem to `jump over' the influence region of the $18/11$ rational surface.
}}
\end{figure}

Figs.~\ref{mean_TE_100_36}-\ref{mean_TE_100_52} show the Transfer Entropy for various magnetic configurations.
Horizontal dashed lines indicate the location of rational surfaces in vacuum (the figures show {\it all} rational surfaces $n/m$ up to the highest value of $n$ and $m$ shown in each figure), while {small circles} indicate the positions of the ECE measurement channels.
The corresponding profiles of the rotational transform, $\iotabar$, are shown in Fig.~\ref{iota}.
The Transfer Entropy graphs are obtained by calculating the $T(\rho,\tau)$, Eq.~\ref{TE}, from the reference ECE channel ($Y$) to the other ECE channels ($X$), at a range of lags, $\tau = k \Delta t$, where $\Delta t$ is the ECE sampling rate. 
The graph is then interpolated linearly on a somewhat finer radial grid for visualization purposes (as the location of the ECE channels is not equidistant in $\rho$).  
This graph is calculated for a number of discharges in a given configuration, and finally an average over these discharges is computed to reduce irrelevant variations (see Section \ref{averaging}).

ECRH power is mainly deposited in the region $|\rho| < 0.2$, so no propagation is expected in that region. 
Also, the Transfer Entropy between the central ECE channel and itself is zero by definition, leading to a horizontal blue streak near $|\rho|=0$ in all graphs, which therefore has no physical meaning.

Outward propagation is visible for $|\rho| > 0.2$, sometimes visible as a diagonal red plume moving outward from the central region.
However, radial propagation is not characterized by a simple, smooth plume but rather displays significant radial structure.

%{[REMOVE: There is a zone of enhanced transfer entropy occurring at $|\rho| \simeq 0.55$ in most graphs, which is probably associated with the zonal flow associated with the (pressure) gradient region located slightly further outward (see Section \ref{discussion}). 
%The barrier at $|\rho| \simeq 0.55$, roughly corresponding to the density `shoulder' seen in Fig.~\ref{TSprof_100_44} (a particle `accumulation zone'), is very systematic and has been associated with the formation of an improved Core Electron Root Confinement (CERC) region in the past~\cite{Estrada:2007}.}

In the region $|\rho| > 0.2$, radial structures are visible that appear to be related to the presence of some low-order rational surfaces, the vacuum locations of which have been indicated in the graphs as horizontal dashed lines.
Although small net currents (typically, with $|I_p| < 1$ kA) may flow in these plasmas, it is expected that such currents only have a minor impact on the rotational transform ($\iotabar$) profile~\cite{Estrada:2002,Velasco:2011}.
%Thus, the rational surfaces in the presence of plasma may be shifted slightly outward with respect to the indicated locations, such that the outward shift decreases with increasing values of $|\rho|$.

It can be seen how the outward radial propagation of the `plume' is slowed at or near specific radial locations, different for each configuration.
The front of the `plume' sometimes shows this slowing down effect, giving it the appearance of a rounded `staircase' (indicated by arrows in some of the graphs). 
In addition, the `plume' is lengthened significantly in the direction of increasing time lag at specific radial locations (as explained in detail in the figure captions).

\clearpage
\subsection{Power modulation experiments}

ECRH power modulation experiments have also been performed, with the goal of analyzing heat transport using standard Fourier techniques~{\cite{Eguilior:2003,Milligen:2016}, although no full iota scan was made}. 
In these experiments, one of the two ECRH systems was operating continuously at $\sim 200$ kW, while the other ECRH system was modulated at low power, with a square waveform.
The line average density was low, $\overline n_e \simeq 0.4 \cdot 10^{19}$ m$^{-3}$.

\begin{figure}\centering
  \includegraphics[trim=0 0 0 0,clip=,width=16cm]{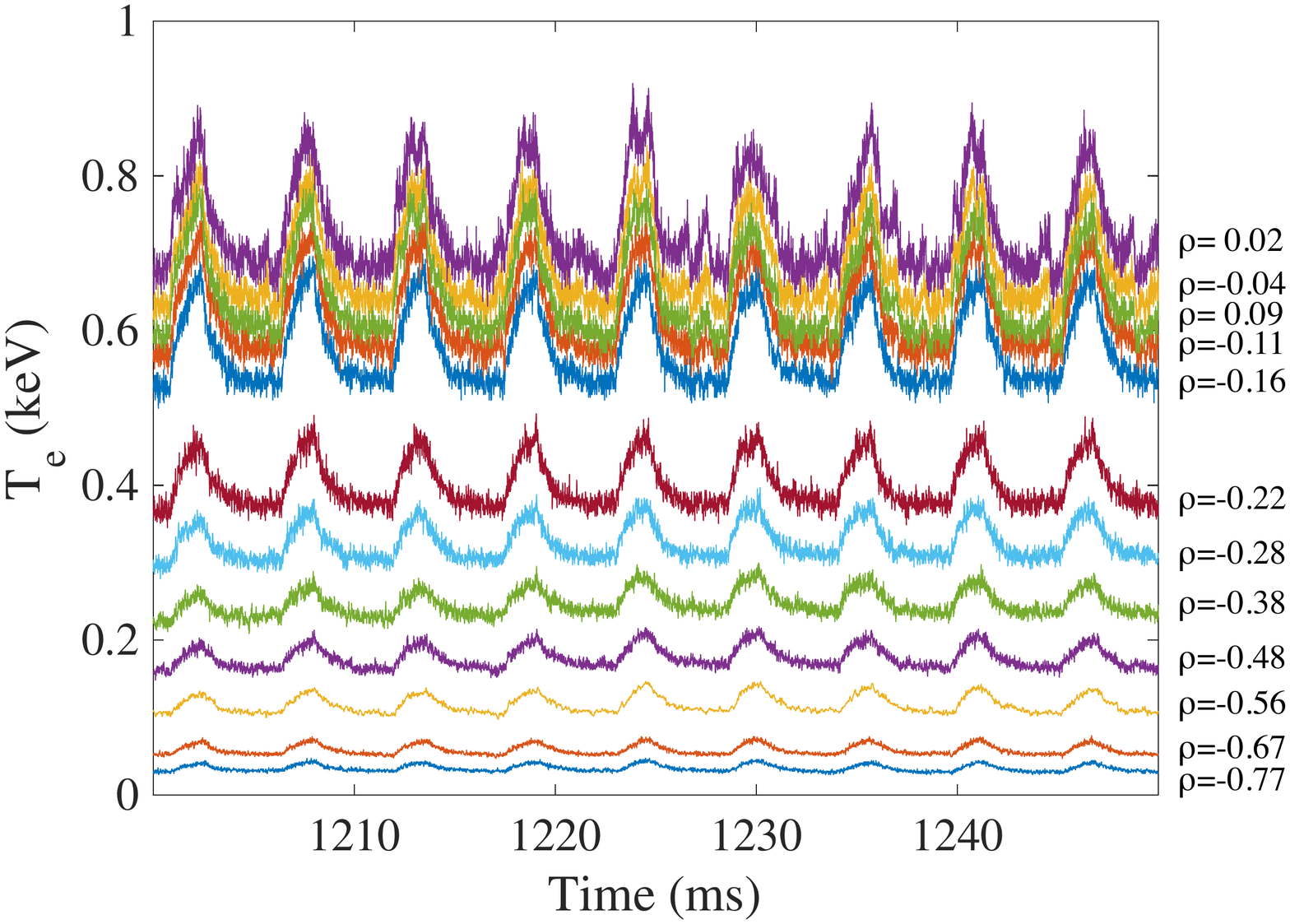}
\caption{\label{modulation_rawdata}{ECE data in a discharge (29748) with modulated central ECRH power. 
Labels indicate the radial position ($\rho = r/a$) of the measurement.}}
\end{figure}

\begin{figure}\centering
  \includegraphics[trim=0 0 0 0,clip=,width=16cm]{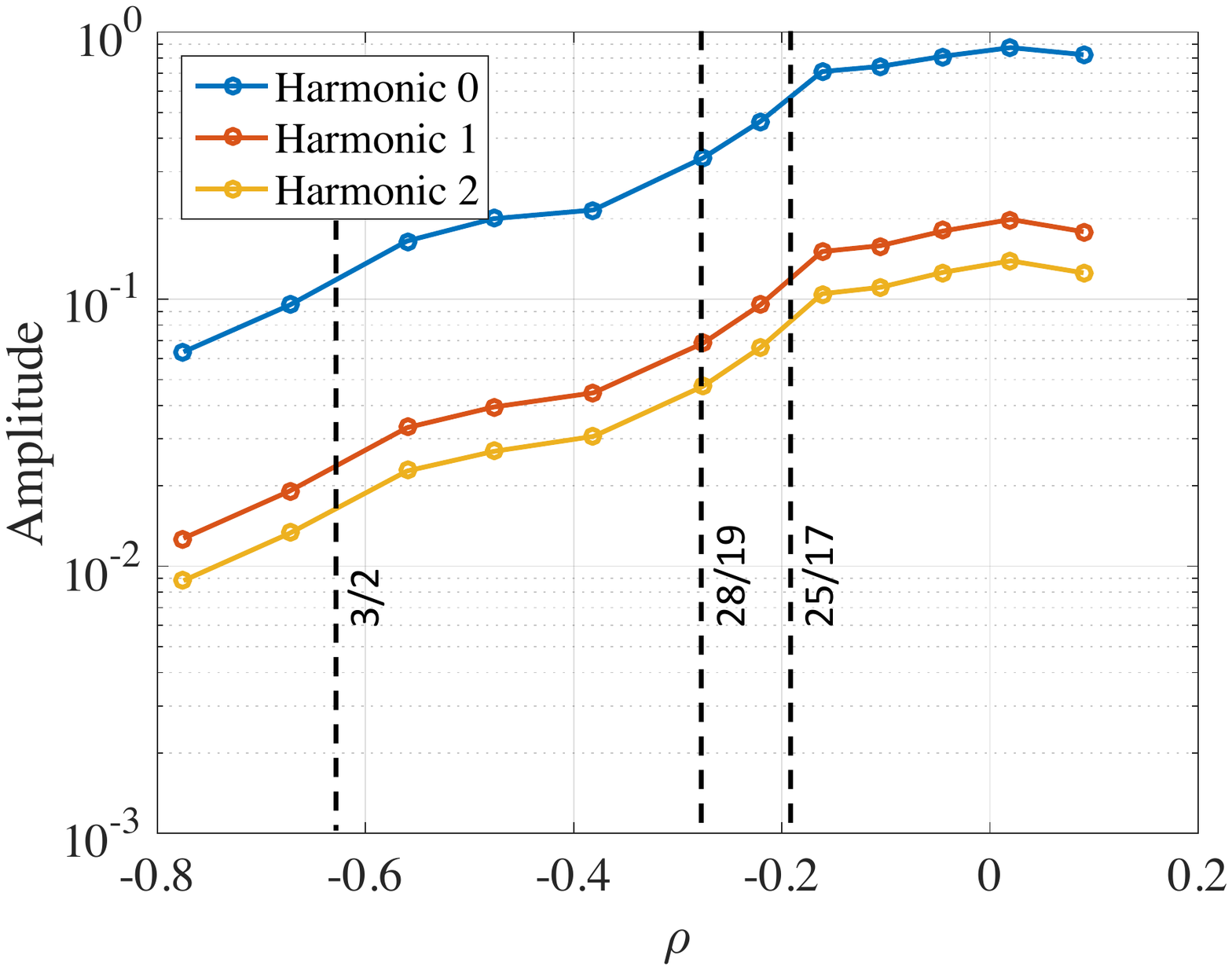}
\caption{\label{modulation_amplitude}{Modulation amplitude at the modulation frequency and the first two harmonics for the discharge shown in Fig.~\ref{modulation_rawdata}.
The positions of a few rational surfaces in vacuum are indicated by vertical dashed lines.}}
\end{figure}

\begin{figure}\centering
  \includegraphics[trim=0 0 0 0,clip=,width=16cm]{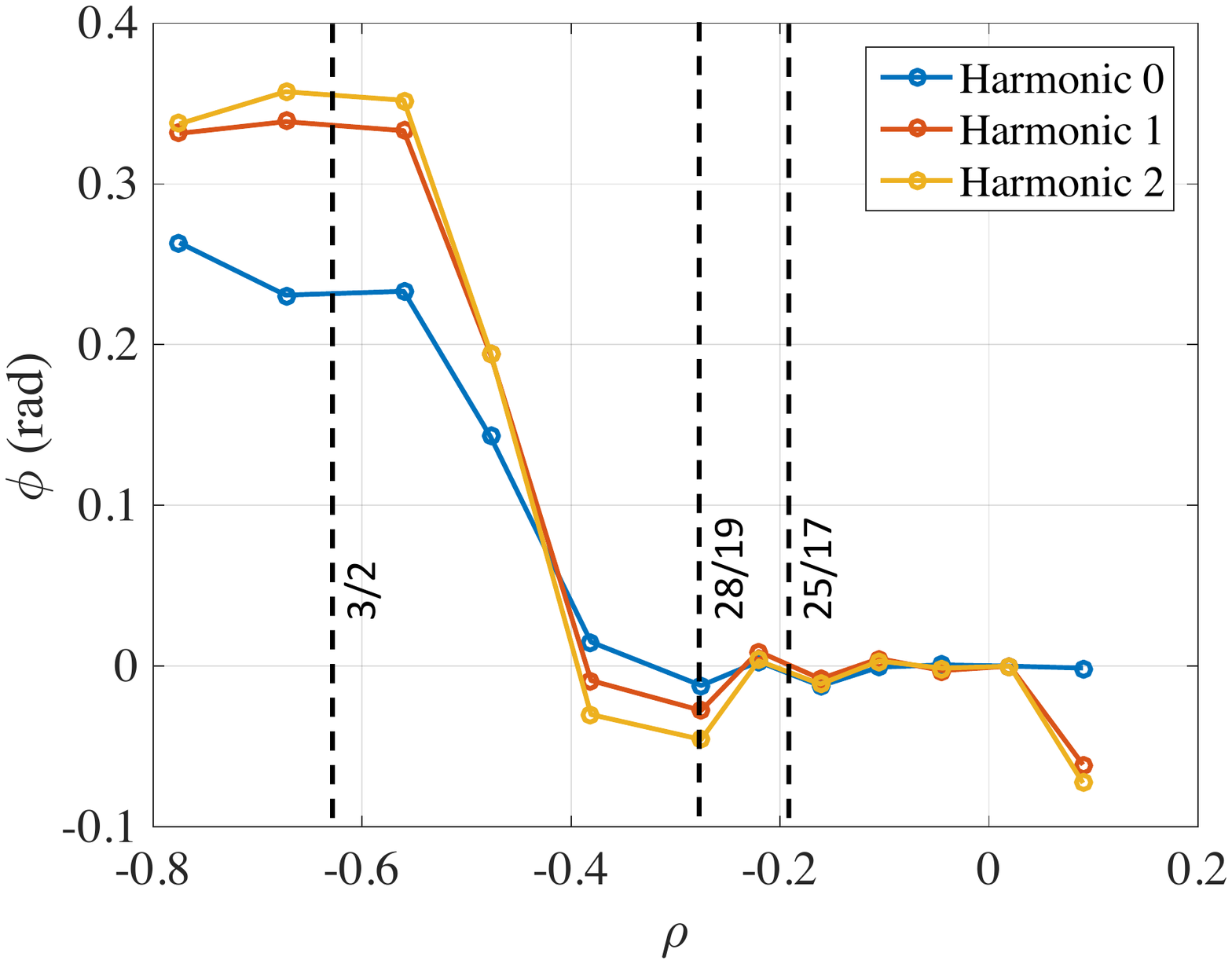}
\caption{\label{modulation_phase}{Modulation phase (relative to the central channel at $\rho \simeq 0.02$) at the modulation frequency and the first two harmonics for the discharge shown in Fig.~\ref{modulation_rawdata} (configuration 100\_36).
The positions of a few rational surfaces in vacuum are indicated by vertical dashed lines.
{Error bars (not shown) are similar to those of Fig.~\ref{modulation_phase2}.}
}}
\end{figure}

Fig.~\ref{modulation_rawdata} shows $T_e$ data obtained in a discharge with configuration 100\_36{, having a low order rational surface (3/2) in the edge region (cf.~Fig.~\ref{iota}), and with} modulated ECRH power (modulation frequency: 180 Hz, duty cycle: 30\%).
Only 50 ms of the available 200 ms of data are shown.
Figs.~\ref{modulation_amplitude} and \ref{modulation_phase} show the amplitude $A$ and relative phase $\phi$ (with regard to the most central ECE channel) of the measurement signals at the modulation frequency and the first two harmonics.
In the power deposition region, around $|\rho| = 0$, $\phi \simeq 0$.
The positions of a few rational surfaces in vacuum are indicated.

The slope of the phase, $d\phi/d\rho$, is inversely proportional to the phase velocity of the outward propagating heat waves~{\cite{Jacchia:1991,Berkel:2014}}.
Note that the slope is particularly high in the range $0.4 < |\rho| < 0.55$, suggesting that the heat waves are slower in that region than elsewhere, likely related to the presence of a barrier associated with the major rational surface ($3/2$) located at $|\rho| \simeq 0.63$ (cf.~Ref.~\cite{Ryter:2006}).
Furthermore, the slope is rather low in the range $0.55 < |\rho| < 0.8$, indicating fast propagation in the area surrounding this major rational surface, possibly indicating propagation around the O-point of a magnetic island~\cite{Spakman:2008}.
 
\begin{figure}\centering
  \includegraphics[trim=50 175 50 25,clip=,width=16cm]{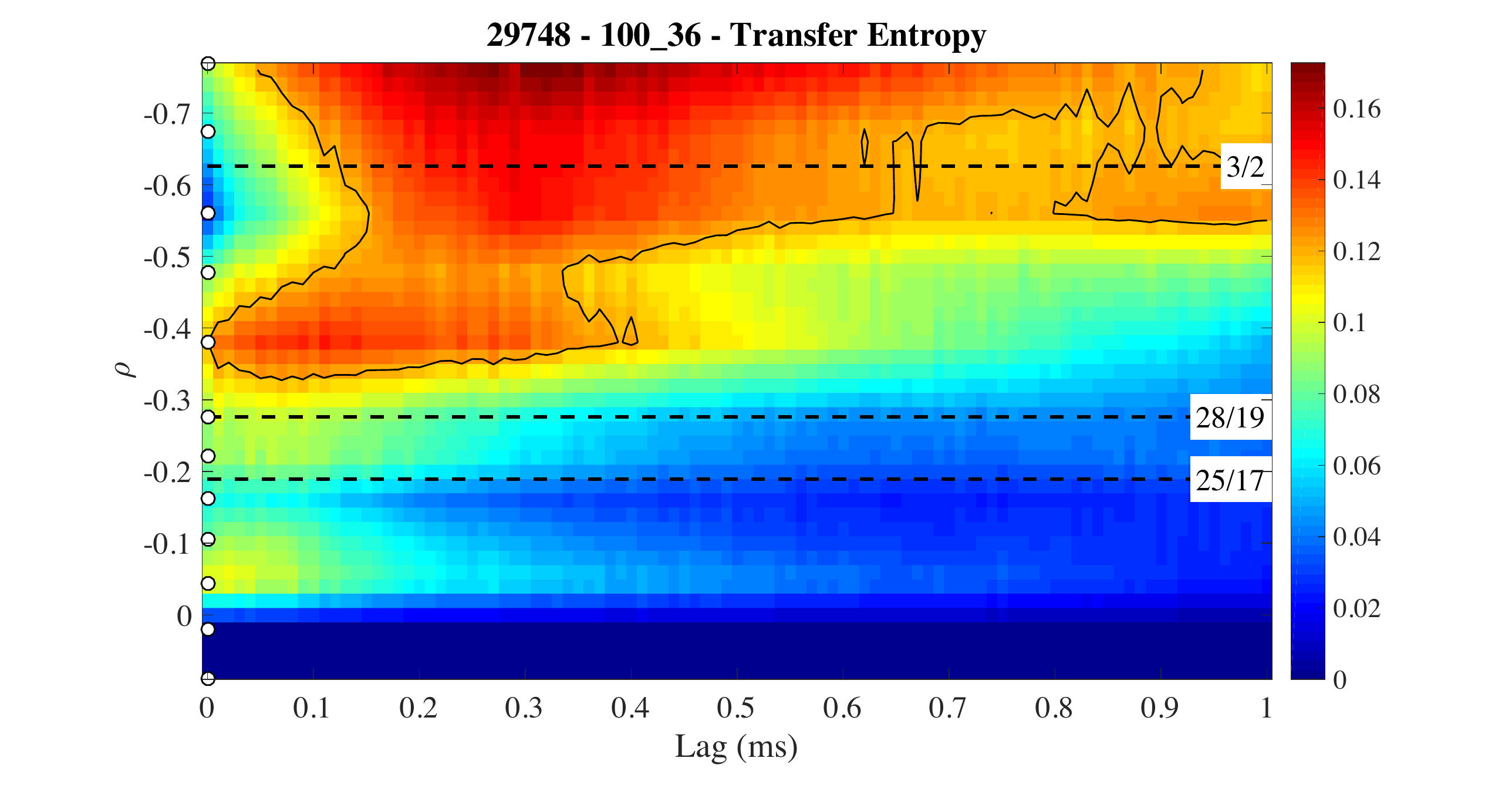}
\caption{\label{modulation_TE}{Transfer Entropy $T(\rho,\tau)$ for the modulated discharge shown in Fig.~\ref{modulation_rawdata}, configuration 100\_36 (see {Fig.~\ref{iota}).}
%Dots indicate the positions of the ECE measurement channels. Horizontal dashed lines indicate the location of rational surfaces in vacuum.
}}
\end{figure}

We then subjected these data to the analysis technique based on the Transfer Entropy.
Fig.~\ref{modulation_TE} shows the result, which should be compared to Fig.~\ref{mean_TE_100_36}.
Particularly interesting is the radial location $|\rho| \simeq 0.55$. 
Whereas in the case of spontaneously generated temperature perturbations, this location acted as a `trapping zone' for the outward propagating temperature perturbations that the outward propagating temperature perturbations were largely unable to cross, here it constitutes only a weak barrier that is crossed easily by the heat waves produced by the modulation, resulting in high values of the Transfer Entropy at locations outside the $3/2$ rational surface ($|\rho| > 0.7$).

Another interesting observation is that propagation appears to slow down inside from $|\rho| \simeq 0.5$, suggesting the existence of a minor transport barrier (possibly associated with the 3/2 rational surface) and low phase velocity (according to Fig.~\ref{modulation_phase}). The latter reinforces the identification of this zone as a transport barrier. 

\begin{figure}\centering
  \includegraphics[trim=0 0 0 0,clip=,width=16cm]{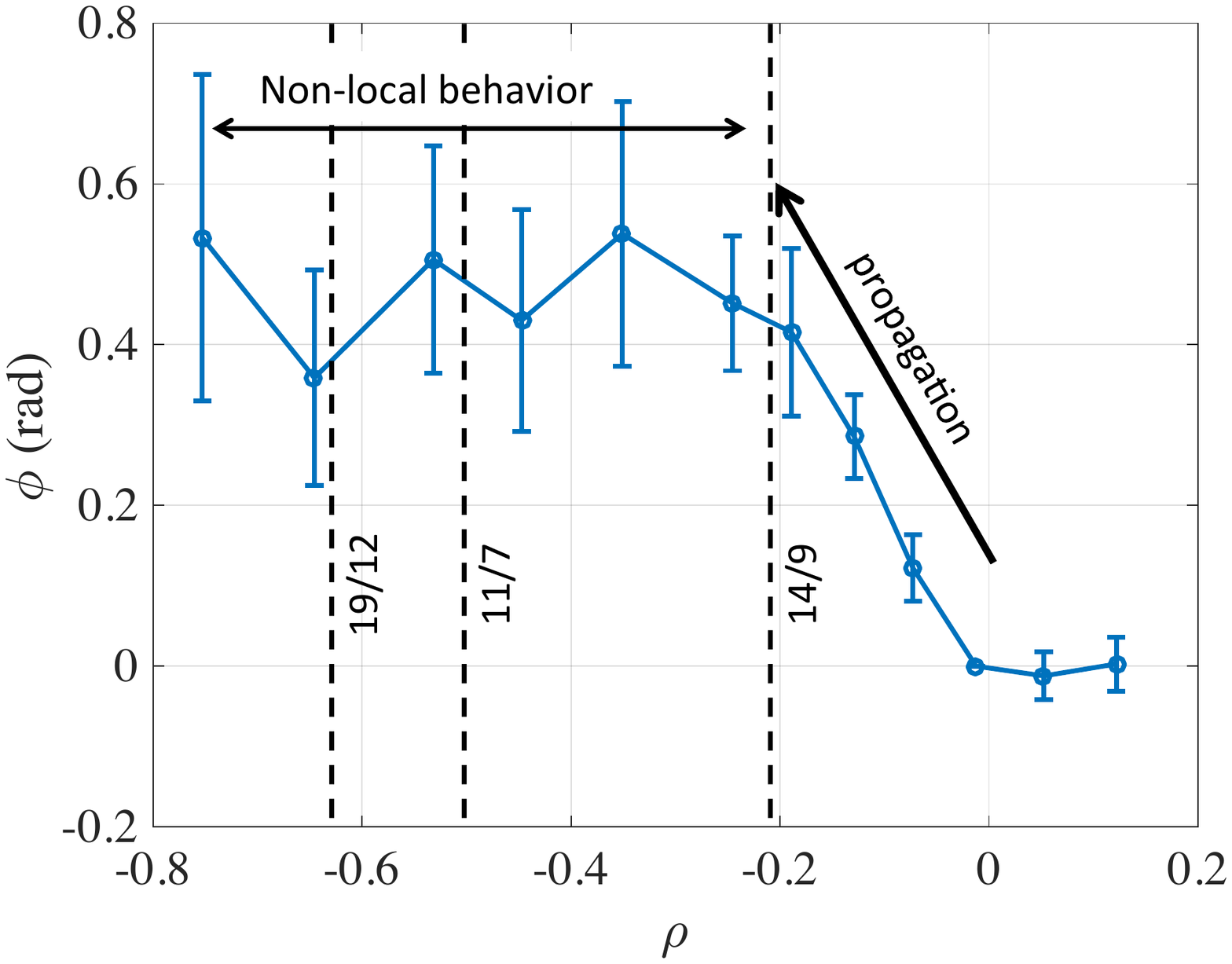}
\caption{\label{modulation_phase2}{Modulation phase (relative to the central channel at $\rho \simeq -0.01$) at the second harmonic of the modulation frequency (configuration {100\_44).}
%The positions of a few rational surfaces in vacuum are indicated by vertical dashed lines.
}}
\end{figure}

\clearpage
Another ECRH modulation case is shown in Fig.~\ref{modulation_phase2}.
Here, the magnetic configuration is 100\_44 (cf.~Fig.~\ref{iota}) and the modulation frequency is 110 Hz. 
The figure shows the phase of the second harmonic of the modulation frequency relative to the central channel at $\rho \simeq -0.01$.
There are clearly two regions: one in which the phase increases gradually with $|\rho|$, indicating propagation, and one in which the phase is approximately constant with $\rho$, suggesting a non-local (simultaneous) response, indicated in the figure with arrows. 

\begin{figure}\centering
  \includegraphics[trim=50 175 50 25,clip=,width=16cm]{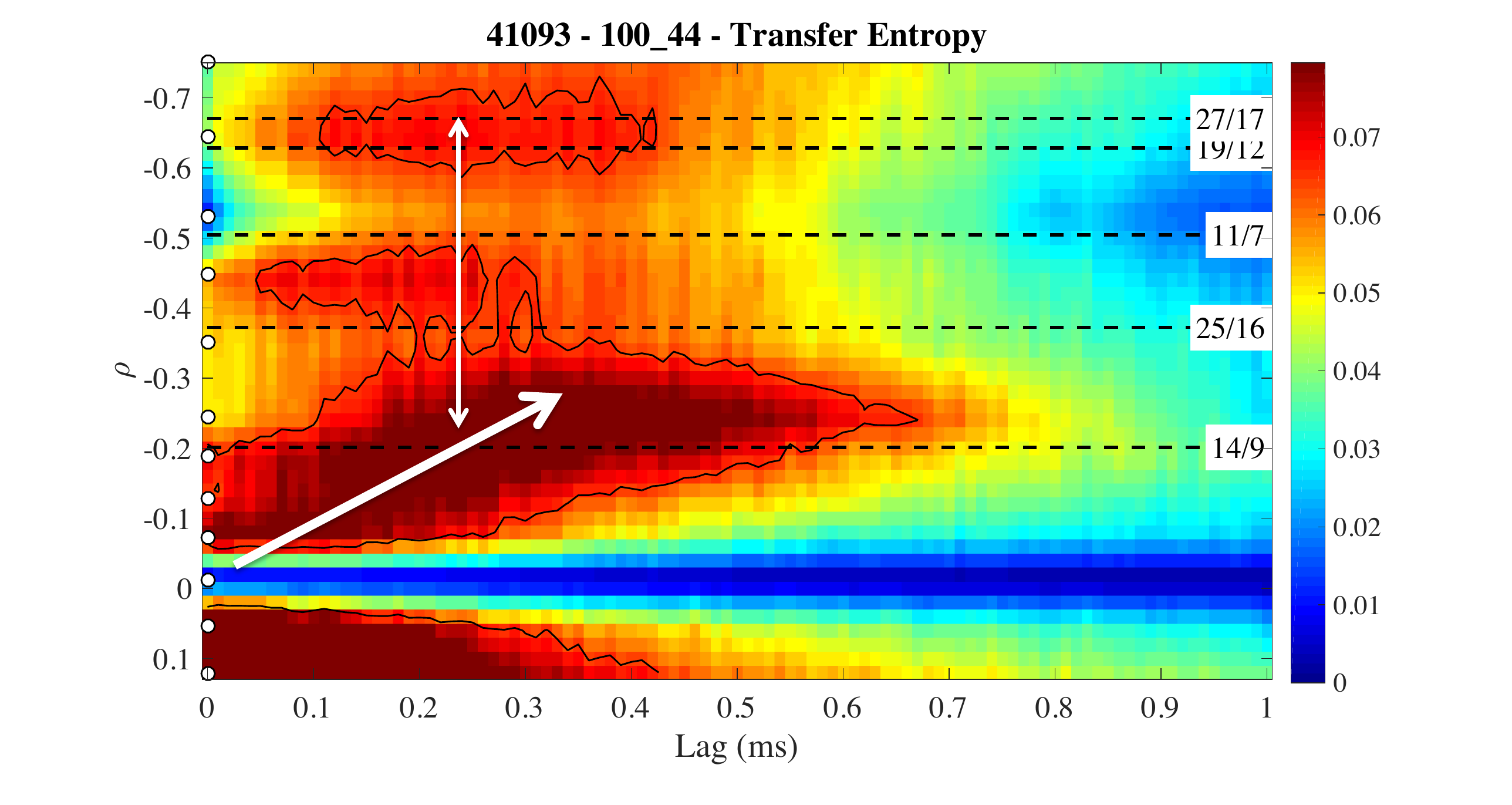}
\caption{\label{modulation_TE2}{Transfer Entropy $T(\rho,\tau)$ for a modulated discharge with configuration 100\_44 (see {Fig.~\ref{iota}).}
%Dots indicate the positions of the ECE measurement channels. Horizontal dashed lines indicate the location of rational surfaces in vacuum.
}}
\end{figure}

The analysis of this discharge using the Transfer Entropy is shown in Fig.~\ref{modulation_TE2}.
This graph likewise seems to suggest the existence of a zone dominated by outward propagation ($|\rho| < 0.3$), and a zone dominated by a simultaneous response ($|\rho| > 0.3$). 
At $|\rho| \simeq 0.25$, a `tail' is seen to develop in the Transfer Entropy, suggesting a `trapping zone', possibly associated with the 14/9 rational surface.
At the time lag the outward propagating pulses (indicated by the thick arrow) cross the 14/9 rational surface, instantaneous responses are detected further outward (specifically, along the two-headed arrow at $|\rho| \simeq 0.45$ and $|\rho| \simeq 0.65$), which we ascribe to mode coupling effects.
Thus, mode coupling is suggested as the explanation for the `non-local' response seen in Fig.~\ref{modulation_phase2}  {(cf.~Section \ref{discussion})}.

%===========================
\clearpage
\section{Resistive Magneto-HydroDynamic model}\label{modeling}

The purpose of the calculations presented here is to aid the interpretation of some of the experimental observations presented in the previous sections. 
Using a resistive MHD model, we study the evolution of heat pulses and the impact of zonal flows on their propagation. 
The presence of low-order rational surfaces leads to the formation of magnetic islands and associated turbulent vortices. 
These, in turn, lead to the generation of zonal flows, and the radial shear of the latter produces what may be called mini-transport barriers~\cite{Carreras:2014,Garcia:2015}.
Zonal flows modify transport and subdiffusive, diffusive or super-diffusive regions may appear, as was studied in earlier work~\cite{Milligen:2016}.
In the present work, we focus on the impact of such regions on the evolution of a heat pulse.
  
The model used is a two-fluid resistive MHD turbulence model which has been used in the past to interpret some of the results from TJ-II experiments~\cite{Garcia:2001}. 
It is based on the Reduced MHD equations \cite{Strauss:1976}, the dominant instability being  pressure gradient driven modes. 
The geometry of the system is a periodic cylinder $\left(r,\theta,z \right)$, where the {averaged} magnetic field line curvature is given by
{
\begin{equation}
\kappa = \frac{r}{R_0} B_0^2V^{\prime \prime},
\end{equation}
}
where $r$ is the radial coordinate, $a$ is the radius of the cylinder, {$R_0$ is an effective major radius and $B_0$ is the toroidal magnetic field.} 
The prime denotes the derivative with respect to toroidal flux, and $V^{\prime}=\int dl/B$ is the specific volume enclosed by a flux surface.
The set of model equations is
\begin{eqnarray}
\frac{\partial \tilde \psi}{\partial t} &=& \nabla_{\parallel}\phi - S \bar \omega_{*e} \left(\frac{T_{eq}}{n_{eq}}\nabla_{\parallel}n + \nabla_{\parallel}T_e \right) + \eta \tilde{J}_z, \nonumber \\
\frac{\partial \tilde U}{\partial t} &=& - v_{\perp} \cdot \nabla U + S^2\nabla_{\parallel}J_z - S^2\frac{\beta_0}{2\varepsilon^2}\kappa \left(\frac{T_{eq}}{n_{eq}} \frac{1}{r} \frac{\partial \tilde n}{\partial \theta} + \frac{1}{r}\frac{\partial \tilde T_e}{\partial \theta} \right) + \mu \nabla_{\perp}^2 \tilde U, \nonumber \\
\frac{\partial \tilde n}{\partial t} &=& - v_{\perp} \cdot \nabla n + \frac{S}{\bar \omega_{ci}}\nabla_{\parallel}J_z + D_{\perp} \nabla_ {\perp}^2 \tilde n, \nonumber \\  % + D_{\parallel} \nabla_ {\parallel}^{(0)2} \tilde n
\frac{\partial \tilde T_e}{\partial t} &=& - v_{\perp} \cdot \nabla T_e + \frac{S}{\bar \omega_{ci}}\frac{T_{eq}}{n_{eq}}\nabla_{\parallel}J_z + \chi_{\perp} \nabla_ {\perp}^2 \tilde T_e + \nabla_{\parallel}\left(\chi_{\parallel}\nabla_{\parallel} Te\right).
\end{eqnarray}
Here, $\psi$ is the poloidal magnetic flux, $v_\perp$ is the perpendicular component of the velocity, $U$ is the toroidal component of the vorticity, $n$ is the density and $T_e$ is the electron temperature. The tildes indicate fluctuations {and the subindices ``eq'' indicate equilibrium profiles}. $\tilde U=\nabla_{\perp}^2\phi/B_0$ where $\phi$ is the electrostatic potential. $J_z=\nabla^2_{\perp}\psi$ is the toroidal current and $\eta$ is the resistivity. The viscosity coefficient is $\mu$, %$D_{\parallel}\left(D_{\perp} \right)$ 
$D_{\perp}$
is the perpendicular density diffusion and $\chi_{\parallel}\left(\chi_{\perp} \right)$ is the parallel (perpendicular) heat  conductivity. $\beta_0$ is the ratio of the plasma pressure $p$ and the magnetic pressure, $B_z^2/2\mu_0$, where $\mu_0$ is the vacuum permeability. 
The resistive time  is $\tau_R=\mu_0 a^2/ \eta \! \left(0 \right)$ where $\eta \! \left(0 \right)$ is the resistivity at the magnetic axis. 
The Alfv\'en time is $\tau_A=R_0\sqrt{\mu_0 m_i n_i}/B_z$ where $m_i$ and $n_i$ are the ion mass and density, respectively. 
The Lundquist number is $S=\tau_R/\tau_A$. 
The inverse aspect ratio is $\varepsilon=a/R_0$. 
The normalized frequencies appearing in the equations are $\bar \omega_{*e}=\tau_A \omega_{*e}$, where $\omega_{*e}=T_e/\left(ea^2B_z \right)$ is the electron diamagnetic frequency; and $\bar \omega_{ci}=\tau_A \omega_{ci}$, where $\omega_{ci}=m_i/\left(eB_z\right)$ is the ion cyclotron frequency.

In the simulations, $S=2 \times 10^5$, $\beta_0=10^{-3}$, $\bar \omega_{*e}=2 \times 10^{-4}$, $\bar \omega_{ci}=500$. The density and temperature values are normalized to the maximum density and temperature at the equilibrium, respectively. 
Lengths are normalized to the minor radius $a$ and times to $\tau_R$.  
An extended explanation of the model can be found in Ref.~\cite{Garcia:2001}.

We will focus on the evolution of the averaged electron temperature which is described by
\begin{equation}
n_{eq}\frac{\partial \left\langle T_e \right\rangle}{\partial t}=-\frac{1}{r}\frac{\partial}{\partial r}\left(r \Gamma_Q \right)+ n_{eq}\frac{1}{r}\frac{\partial}{\partial r} \left( r \chi_{0 \perp}\frac{\partial \left\langle T_e \right\rangle}{\partial r}\right) +S_Q,
\end{equation}
where the heat flux $\Gamma_Q$ is
\begin{equation}
\Gamma_Q=n_{eq}\left\langle \tilde v_r \tilde T_e\right\rangle - \frac{S}{\bar{\omega}_{ci}}T_{eq}\left\langle \tilde B_r \tilde J_z \right\rangle - n_{eq} \left\langle \tilde B_r \chi_{\parallel} \nabla_{\parallel} T_e \right\rangle .
\end{equation}
The angular brackets indicate poloidal and toroidal angle average. 
$v_r$ and $B_r$ are the radial velocity and magnetic field, respectively. 
The source $S_Q$ is chosen so as to keep the integrated electron temperature constant. 

\begin{figure} \centering
 \includegraphics[scale=0.3]{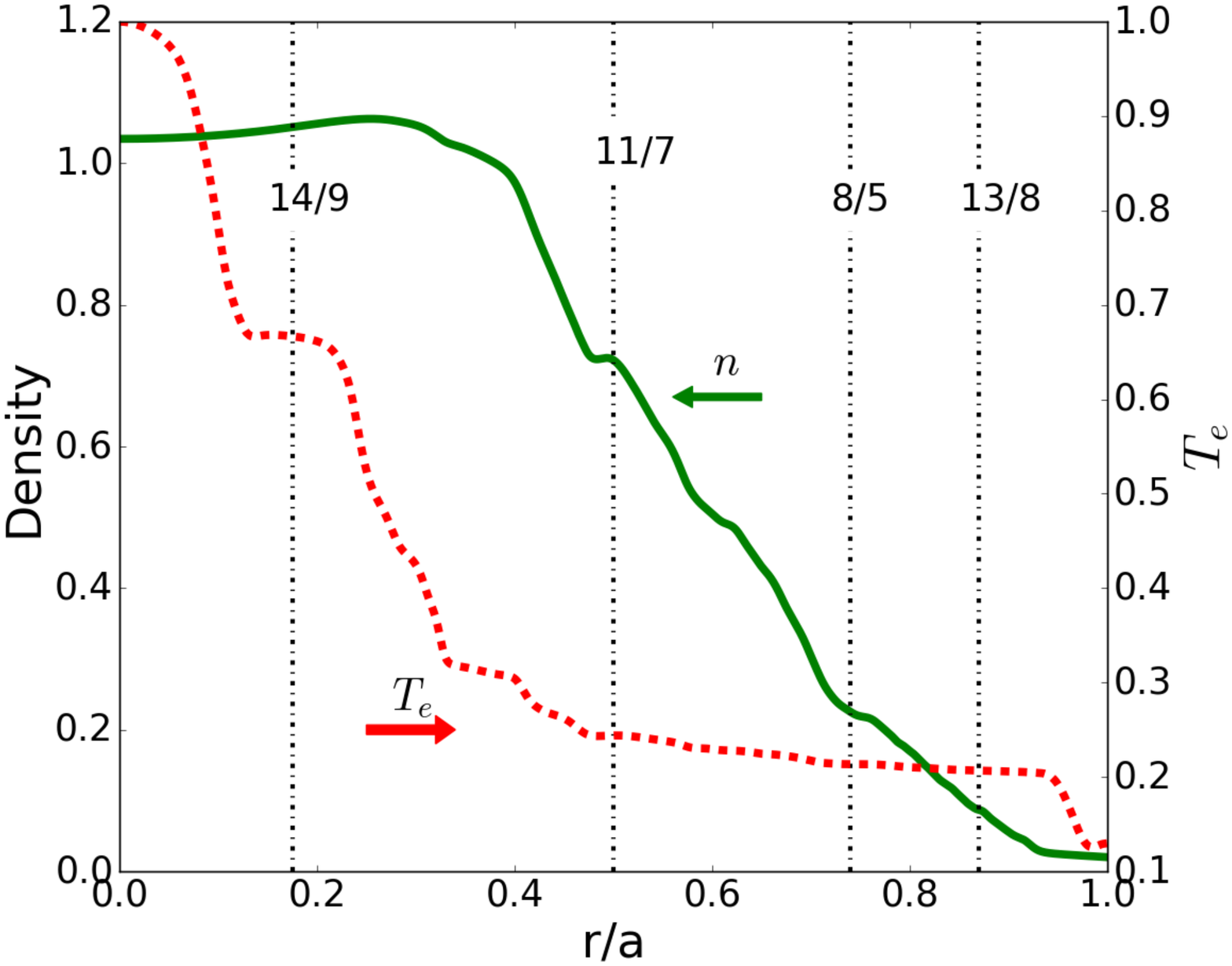}
 \caption{Density and electron temperature profiles used in the simulations. Vertical dashed lines represent main rational surfaces.} 
 \label{fig:steady_state_profiles}
 \end{figure}

We have used the rotational transform profile of configuration  100\_44, as shown in Fig.~\ref{iota}. 
Figure \ref{fig:steady_state_profiles} displays the initial density and electron temperature profiles used in the simulations.
They are obtained by self-consistently evolving the equations until a steady state is obtained.
These profiles are characterized by flat regions around low-order rational surfaces, which is the result of plasma self-organization, as explained elsewhere~\cite{Ichiguchi:2011}.
  
To study electron temperature transport in the framework of this model, a periodic heat perturbation is introduced in the inner region of the plasma. After a short transient, the evolution produced by the periodic pulses is followed over time. The pulses are instantaneous Gaussian perturbations with a maximum $\Delta T_e=0.1$, width $\sigma=0.007$, periodicity $t=5 \times 10^{-4} \tau_R$ and located at $r/a=0.15$. 

\begin{figure}
	\centering
 \includegraphics[scale=0.3]{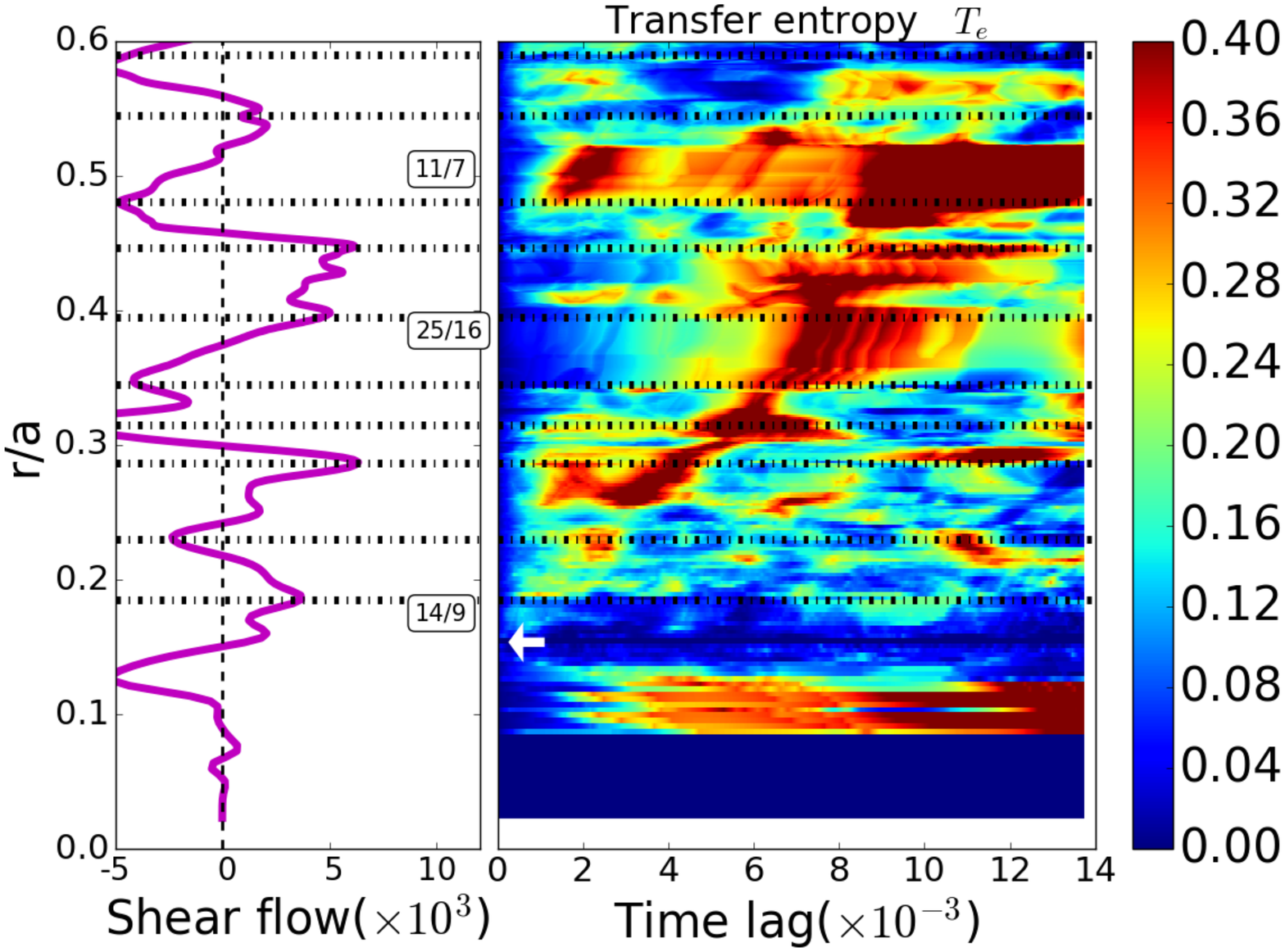}
 \caption{The left panel displays the flow shear $\left( d\! \left\langle \! V_{\theta} \right\rangle \! / \! dr \right)$ at the initial time.
 The right panel shows the Transfer Entropy of the periodic {pulses.} 
%Dashed horizontal lines indicate flow shear maxima and minima. The location of main rational surfaces is displayed by the text boxes. The white arrow points the position of the reference signal and the inital location of pulses.
 } 
 \label{fig:periodic_pulses}
 \end{figure}

The Transfer Entropy method was then applied to these numerical data. The left panel in Fig.~\ref{fig:periodic_pulses} displays the flow shear $\left( d\! \left\langle v_{\theta} \right\rangle \! / \! dr \right)$ (where $\left\langle v_{\theta} \right\rangle$ is the poloidal and toroidal average of the instantaneous poloidal velocity $v_{\theta}$) at the initial time. Although this averaged poloidal velocity does evolve in time, its variation over the time window studied is small enough to ignore.
{In this and the following figures, horizontal dashed lines indicate local flow shear maxima and minima to facilitate interpretation. 
Text boxes indicate the position of the main rational surfaces, and the white arrow points the position of the reference signal and the initial location of pulses.}
One observes that the position of the rational surfaces is not always a good predictor of the exact location of flow shear maxima and minima and the corresponding transport barriers.

The right panel displays the Transfer Entropy of the electron temperature data, averaged over the toroidal and poloidal angles, which is similar to, though clearer than, the Transfer Entropy  calculated from $T_e$ at a single poloidal and toroidal angle.
The white arrow in $r/a=0.15$ indicates the origin of the pulses and the location of the reference signal.
Clearly, the Transfer Entropy shows radial propagation from $r/a \approx 0.2$ to $r/a \approx 0.5$. 
In this radial range, the amplitude of the Transfer Entropy varies, and these variations match the flow shear maxima and minima. 
In addition, non-local effects can be observed around $r/a \approx 0.1$ and $r/a \approx 0.5$.

\begin{figure}
	\centering
 \includegraphics[scale=0.3]{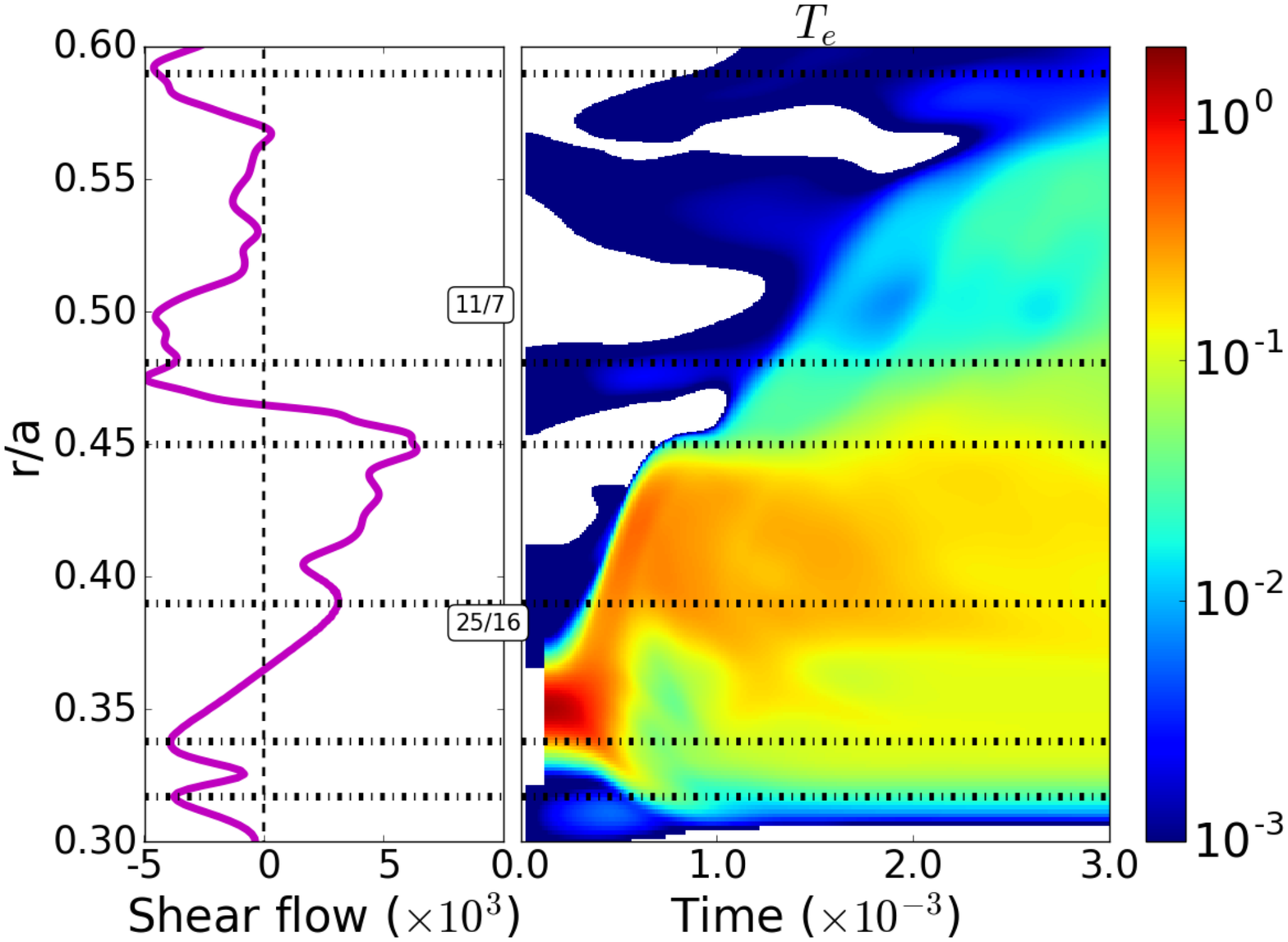}
 \caption{The left panel displays the flow shear $\left( d\! \left\langle \! V_{\theta} \right\rangle \! / \! dr \right)$ at the time of the perturbation.
 The right panel shows the evolution of the electron temperature {perturbation}. 
 %Dashed horizontal lines indicate flow shear maxima and minima. The location of main rational surfaces is displayed by the text boxes.
 } 
 \label{fig:pulse_0.35_Te}
 \end{figure}

It is also interesting to study the evolution of a single heat pulse. In a different numerical simulation without  periodic pulses, an instantaneous heat perturbation is introduced in the inner region of the plasma, after which the evolution produced by the model is followed over time. 
The initial perturbation is Gaussian with peak value $\Delta T_e=1.5$ and a narrow shape, with width $\sigma=0.007$. 
The result of the numerical experiment is shown in Fig.~\ref{fig:pulse_0.35_Te}. 
The left panel displays the flow shear $\left( d\! \left\langle v_{\theta} \right\rangle \! / \! dr \right)$ immediately before the pulse.
The right panel displays the evolution of the electron temperature. The colors indicate the temperature difference between the initial $T_e$ profile (immediately before the pulse) and evolving values of $T_e$. The horizontal dashed lines indicate flow shear maxima and minima.

The right panel of Fig.~\ref{fig:pulse_0.35_Te} shows that the pulse spreads similar to what would occur in a diffusion-type system. 
However, more complex behavior is also visible:
namely, the temperature perturbation propagates from $r/a\approx 0.35$ to $r/a\approx 0.45$, where it slows down, thus creating a `step' in the pulse front. 
The latter position coincides with a maximum of flow shear, explaining why radial propagation is slowed down in this particular location. 
Another such `step' can be observed at $r/a\approx0.48$, coinciding with another maximum of (absolute)  flow  shear.

\begin{figure} \centering
 \includegraphics[scale=0.3]{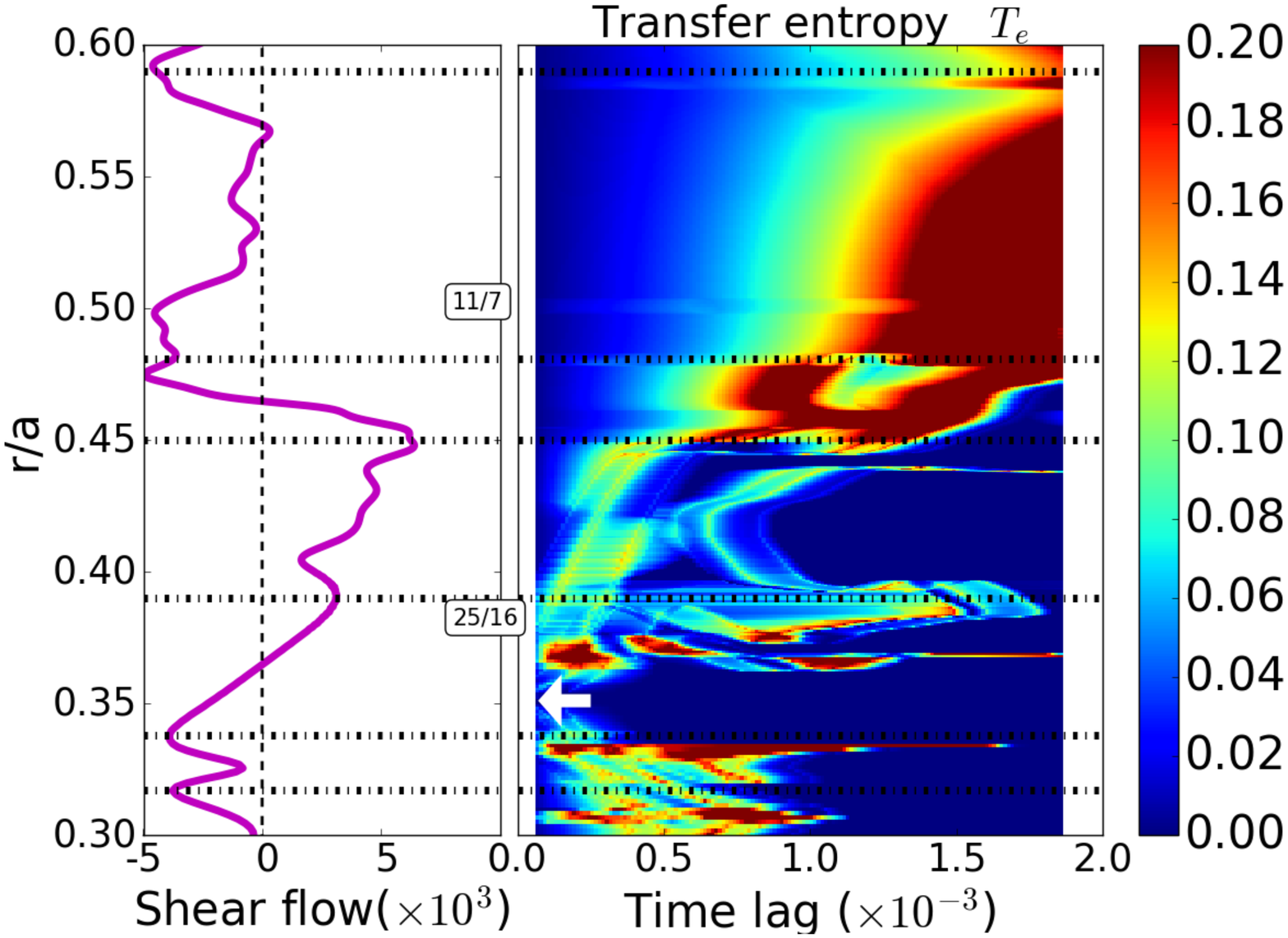}
 \caption{The left panel displays the flow shear at the time of the perturbation. 
The right panel shows the Transfer Entropy calculated for the pulse shown in {Fig.~\ref{fig:pulse_0.35_Te}.}
%The white arrow indicates the position of the reference signal. Dashed horizontal lines indicate flow shear maxima and minima.
} 
 \label{fig:pulse_0.35_entropy}
\end{figure}

Next, we apply the Transfer Entropy method to these numerical data. 
The result is shown in Fig.~\ref{fig:pulse_0.35_entropy}. 
As before, the left panel displays the flow shear immediately before the perturbation. 
The right panel displays the Transfer Entropy, calculated using the time evolution of the temperature at $r/a=0.35$ as a reference (white arrow). 
Again, one observes radial propagation, slowed down at $r/a=0.45$, coinciding with the maximum flow shear. 
Locations further outward, up to $r/a = 0.55$, correspond to ever larger time lags, as expected from pulse propagation. 
High values of the Transfer Entropy occurring at positions $r/a>0.55$ are probably due to mode coupling effects.

\begin{figure}
	\centering
 \includegraphics[scale=0.3]{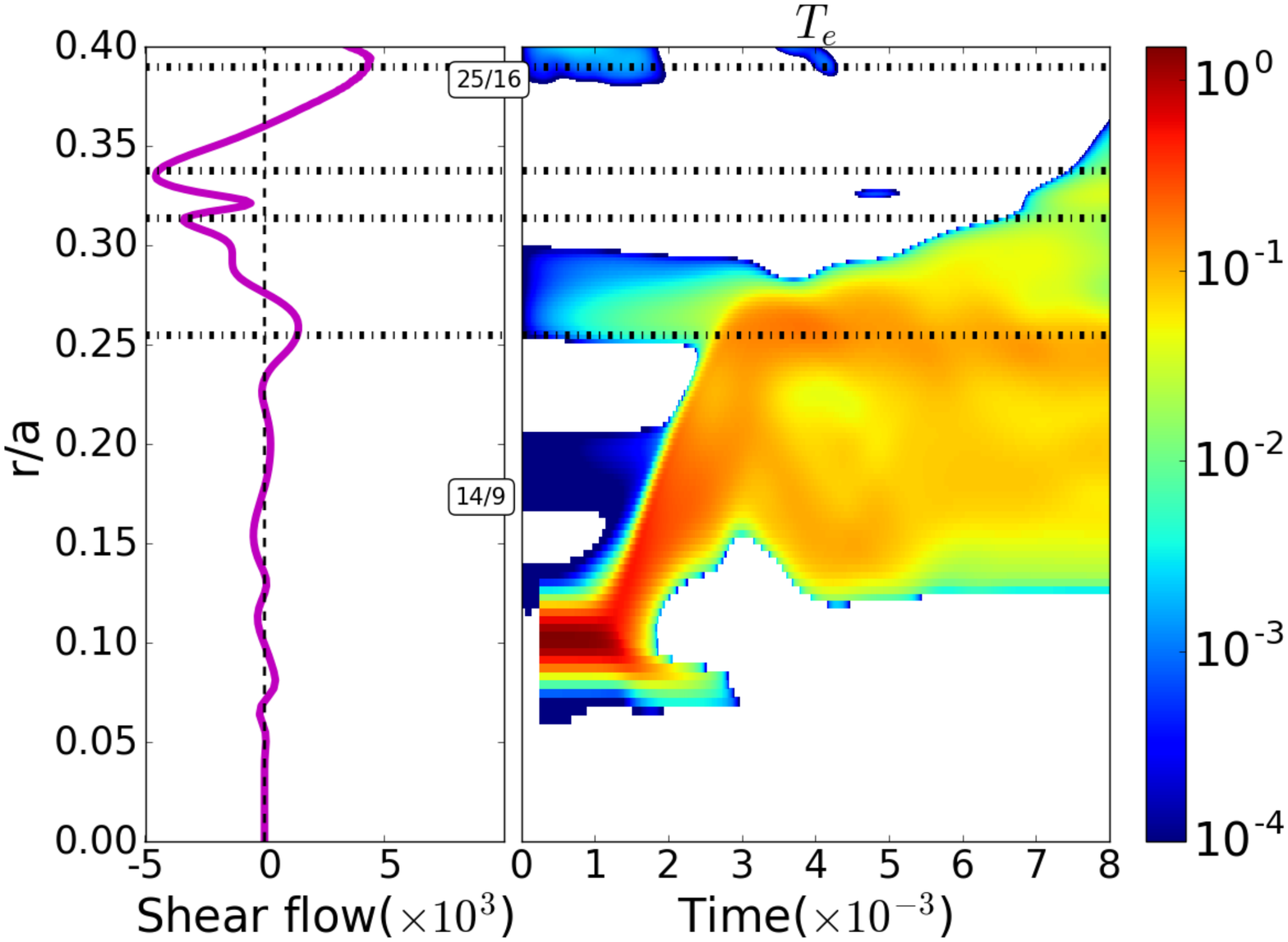}
 \caption{The left panel  displays the flow  shear $\left( d\! \left\langle \! V_{\theta} \right\rangle \! / \! dr \right)$ at the time of the perturbation. 
 The right panel displays the evolution of the electron temperature {perturbation.} 
%Dashed horizontal lines indicate flow shear maxima and minima.
} 
 \label{fig:pulse_0.10_Te}
 \end{figure}

Next, we consider an electron temperature perturbation in a different region for comparison. 
The parameters of the perturbation are the same as before: peak value $\Delta T_e=1.5$ and width $\sigma=0.007$. 
However, the perturbation is now introduced in the core region, $r/a=0.1$, chosen because the average poloidal velocity is almost constant in this region, leading to very small flow shear. 
Figure \ref{fig:pulse_0.10_Te} displays the evolution of the perturbation. 
As before, the left panel displays the flow  shear at the initial time, while the right panel displays the evolution of the temperature perturbation. 
This case shows how the perturbation propagates outward in the radial direction from $r/a\simeq 0.1$ to $r/a\simeq 0.27$, where it slows down. 
Note that the latter position coincides with an increase of flow shear. 

\begin{figure}
	\centering
 \includegraphics[scale=0.3]{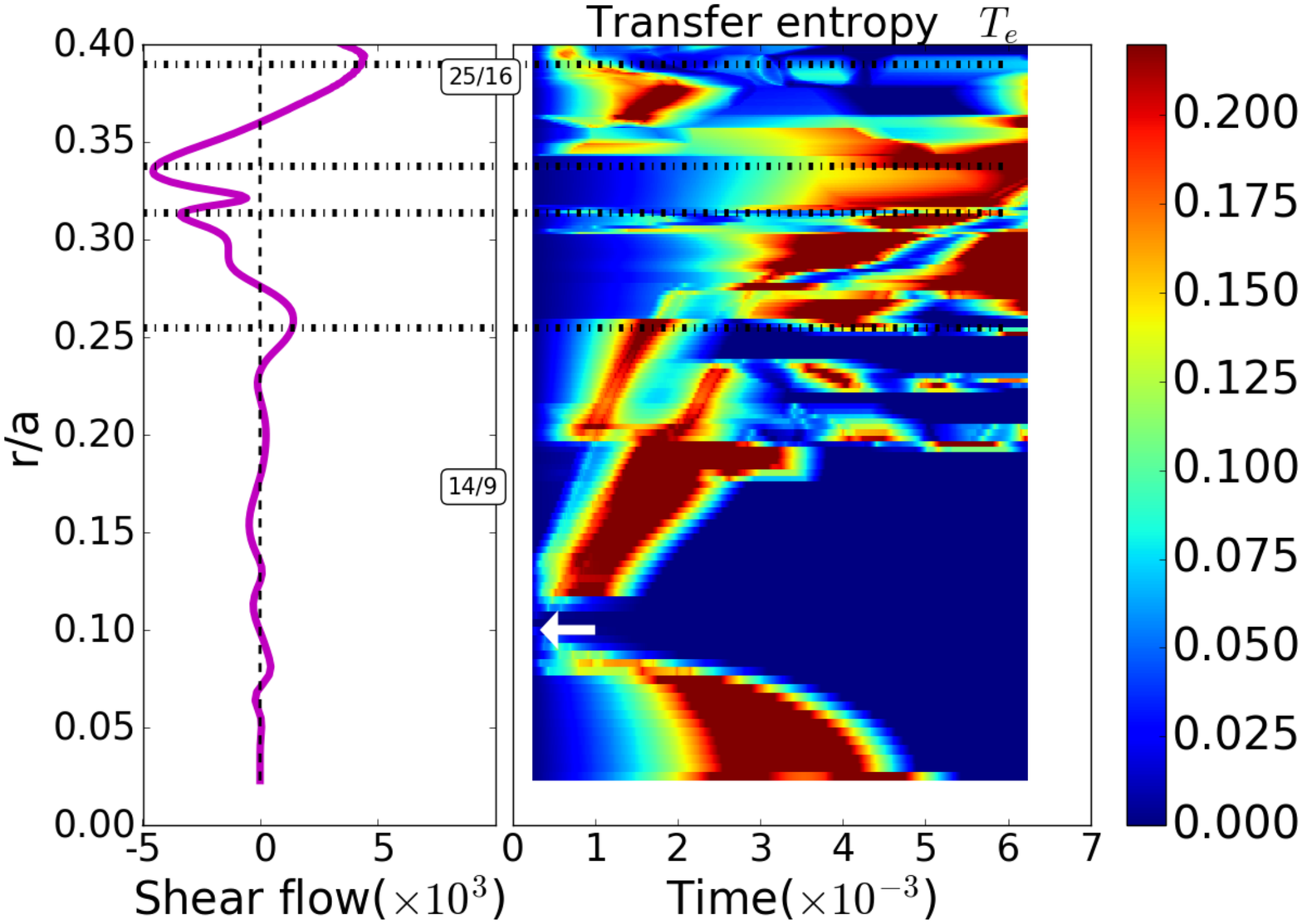}
 \caption{The left panel displays the  flow shear at the time of the perturbation. 
The right panel shows the Transfer Entropy calculated for the pulse shown in {Fig.~\ref{fig:pulse_0.10_Te}.}
%The white arrow indicates the position of the reference signal. Dashed horizontal lines indicate flow shear maxima and minima.
} 
 \label{fig:pulse_0.10_entropy}
 \end{figure}
 
Fig.~\ref{fig:pulse_0.10_entropy} shows the Transfer Entropy for this case. 
The position of the reference signal, at $r/a=0.1$ (the location of the perturbation) is  indicated by a white arrow.
The diagonal structures emanating from the initial position clearly indicate radial propagation. 
This is particularly clear in the range $0.12 < r/a < 0.20$ (first branch).
There is a second branch of propagation occurring in the range $0.20 < r/a < 0.26$ (at time lags below 0.002), that occurs nearly simultaneously with the first branch, suggesting this may be related to a mode coupling effect around the $14/9$ rational surface.
Then, at $r/a \simeq 0.27$, the time lags increase suddenly, suggesting the flow shear (left panel) causes the propagation to slow down.

In both cases analyzed here, electron temperature perturbations propagate radially outwards, similar to what is observed in the experiments.
In addition, one observes that the propagating perturbations are slowed down in regions of increased flow shear, associated with low-order rational surfaces and magnetic islands. 
The propagation front is characterized by a similar `staircase' as observed in the experiment.
Furthermore, the Transfer Entropy sometimes appears to `jump' radially over certain regions (i.e., displaying large values at the same time lag but at very different radial positions), again similar to the experimental observations.
The `jumping' behavior is not evident from the evolution of the temperature perturbation, which is largely continuous; 
the Transfer Entropy therefore reveals underlying behavior that is not accessible using standard analysis techniques.

%===========================
\clearpage
\section{Discussion and conclusions}\label{discussion}

Previous work has suggested that ITBs are often associated with rational surfaces (see the references provided in Section \ref{introduction}).
These ideas also have found some support from simulations~\cite{Thyagaraja:2000,Itoh:2007}.
In spite of the long list of papers suggesting that rational surfaces must have an impact on heat transport, and some intriguing experimental results~\cite{Schilham:2001}, little direct evidence has {emerged}.

This work presents convincing experimental evidence for the influence of low order rational surfaces on heat transport.
This evidence is based on several ingredients:
First, measurements are performed in a stellarator with nearly full external control of the magnetic configuration and the location of the rational surfaces.
Second, use is made of the generation of spontaneous temperature perturbations due to core ECRH heating. 
These perturbations may be generated as a consequence of the presence of rational surfaces in the power deposition region~\cite{Estrada:2002} along with the generation of fast electrons~\cite{Garcia:2016}.
Third, we have applied an advanced analysis technique (the Transfer Entropy) to spatially localized ECE data.
Fourth, we successfully used the technique of  averaging over similar discharges to reduce statistical variation and highlight configuration-dependent features.

The Transfer Entropy is not a correlation or conditional average, and hence requires a different interpretation. 
It detects the `causal impact' of one signal ($Y$) on another one ($X$), at a given time lag $k$.
This `causal impact' is unrelated to signal amplitude or sign, but rather to the number of additional bits of signal $X$ that can be predicted using signal $Y$.
That makes the technique ideally suited to detect the propagation of perturbations:
it is rather robust and independent from irrelevant details such as signal amplitude and the shape of the propagating temperature perturbations, merely focusing on `information transfer' (see Section \ref{transferentropy}).
In addition, it is directional and therefore allows studying outward propagating events while ignoring inward propagating events, resulting in clearer results.

The radial transport was seen to exhibit a number of specific features: (1) radial regions where outward propagation slows down, connected to (2) radial regions where the Transfer Entropy graphs develop `tails' towards long lags; and (3) apparently disconnected `jumps' over certain radial regions, so that regions further out respond at similar lag values as regions further in.

The observed `stepwise' propagation of the temperature perturbations (associated with points 1 and 2) can be understood from the impact of flow shear regions, associated with rational surfaces, on transport, as discussed in earlier work~\cite{Milligen:2016}.
As a consequence, radial propagation may be slowed down near rational surfaces (as also observed on Alcator C-Mod~\cite{Wukitch:2002}),
leading to `stick-slip' behavior and a characteristic `staircase' shape of the propagation front and `tails' developing in the Transfer Entropy plots at long time lags.

On the other hand, perturbations are also observed to `jump over' certain radial regions (point 3), strongly suggesting a `non-local' component of transport. 
{Based on the simulations made with the resistive MHD model, we suggest that this effect is related to resonances occurring between MHD modes located at different rational surfaces. The resulting `instantaneous, non-local' causal influence chains occurring in this model have been studied in previous work~\cite{Milligen:2016c}.
Rapid transport along the boundaries of island O-point regions may also be involved~\cite{Spakman:2008}.}

It is interesting to note that these effects, mostly of magnetic origin, are detected using ECE measurements: this circumstance opens the possibility of studying transport effects due to magnetic perturbations inside hot fusion plasmas, using the techniques employed in this work.

We also studied the propagation of temperature perturbations in a Magneto-HydroDynamic model. The periodic pulses (Fig.~\ref{fig:periodic_pulses}) and the two single pulses studied (Fig.~\ref{fig:pulse_0.35_Te}-\ref{fig:pulse_0.10_entropy}) displayed similar behavior as the experiment. 
The heat pulses propagate radially outwards, but are slowed down at locations with significant flow shear, associated with{, but not located at,} rational surfaces. 
As a result, one observes a similar `staircase' as in the experimental results.
Likewise, the non-local `jumping' effect was also observed in the simulation.
{In general, the spatio-temporal behavior of the fluctuations was found to be rather complex; for example, not only did the local flow shear affect the propagating heat pulses, but the reverse also occurred.}

In addition, we performed ECRH modulation experiments. 
Data from modulated discharges have the advantage that they can be analyzed using standard Fourier-based propagation techniques, facilitating a direct comparison with the technique used here.
As reported above, in configuration 100\_36, the radial propagation (from the Transfer Entropy) was found to slow down towards $|\rho| \simeq 0.5$, which matches the location of relatively slow propagation observed using Fourier techniques.
We conclude that this zone corresponds to a  minor transport barrier.
On the other hand, `non-local' transport is observed in the Transfer Entropy for $|\rho | > 0.55$, corresponding to very fast propagation as deduced from the Fourier analysis (nearly constant phase $\phi(\rho)$), suggesting mode coupling effects occurring around the 3/2 rational surface; or otherwise, rapid transport across the island O-point.
In the other configuration studied, 100\_44, again two zones could be distinguished according to the response of the relative modulation phase: a propagation and a `non-local' response region. The Transfer Entropy confirmed this interpretation and it was suggested that heat perturbations crossing the 14/9 rational surface trigger the mode coupling events that lead to this `non-local' response. 
This second configuration also allows a comparison with the model simulations. 
Indeed, Fig.~\ref{fig:pulse_0.10_entropy} clearly shows that the propagating heat pulse crossing the 14/9 rational surface triggers a simultaneous response at the 25/16 rational surface, similar to the experimental observations.
This agreement between modulation results (which can be interpreted in terms of an effective heat transport coefficient) and results from the Transfer Entropy implies that the results reported here may be quite significant for heat transport in TJ-II.

{The $T_e$ oscillations produced by ECRH modulation are quite large as compared to the spontaneous temperature perturbations (cf.~Figs.~\ref{rawdata} and \ref{modulation_rawdata}). Consequently, local plasma conditions, including zonal flows and transport barriers, are modified or affected by the modulation and it should not be expected that the Transfer Entropy picture is the same in these cases. 
Indeed, a comparison between Figs.~\ref{mean_TE_100_36} and \ref{modulation_TE}, on the one hand, and between Figs.~\ref{mean_TE_100_44} and \ref{modulation_TE2}, on the other, shows that the modulation cases differ significantly from the steady state cases.}

In a recent publication, it was suggested that turbulence may self-organize into a state with a succession of localized shear layers, the so-called `$E\times B$ staircase'~\cite{Dif-Pradalier:2015}. This `$E\times B$ staircase' is unrelated to rational surfaces, whereas the local barriers detected here seem to be clearly associated with rational surfaces, with the possible exception of the barrier at $|\rho| \simeq 0.55$ that is almost always present, regardless of the location of the rational surfaces. Thus, the barrier at $|\rho| \simeq 0.55$ may indeed be produced by the mechanism suggested in Ref.~\cite{Dif-Pradalier:2015}, but we note that the other barriers in the range $0.2 < |\rho| < 0.55$ are more likely of the type described in Ref.~\cite{Milligen:2016}: subdiffusive regions and flow shear layers associated with rational surfaces.
It is interesting to note that this would mean that both types of transport barriers can be studied in significant detail in TJ-II plasmas.

The radial propagation of heat perturbations revealed by the study of the Transfer Entropy does not look smooth or diffusive. 
It is therefore not easy to deduce a clear radial propagation speed from these results.
Its order of magnitude, $a/2 \simeq 10$ cm in $0.5-1$ ms, or $100-200$ m/s, is consistent with observations reported in earlier work~\cite{Milligen:2002}.
{Considering that the typical confinement time of TJ-II is $\tau_E = 2-6$ ms~\cite{Ascasibar:2005}, one might expect radial propagation velocities of $v_E \simeq a/\tau_E \simeq 35-100$ ms, of the same order as the mentioned propagation speed.}
Nevertheless, it seems inappropriate to attempt to force the observed behavior into the rigid framework of a transport model based merely on convection and diffusion.
Clearly, the situation is more complex.

Similar careful ECE measurements in other stellarators with full external control of the magnetic configuration, such as W7-X~\cite{Koenig:2015}, should allow a confirmation of these results using the same analysis techniques.

%===========================
\section*{Acknowledgements}
{The authors would like the TJ-II team for its continued support.}  
Research sponsored in part by the Ministerio de Econom\'ia y Competitividad of Spain under project 
Nrs.%ENE2012-30832, % I. Calvo
~ENE2015-68265-P % L. García
and ENE2015-68206-P. % B. v. Milligen
This work has been carried out within the framework of the EUROfusion Consortium and has received funding from the Euratom research and training programme 2014-2018 under grant agreement No 633053. 
The views and opinions expressed herein do not necessarily reflect those of the European Commission.

%===========================
\clearpage
\section*{References}

%\bibliography{/Users/milligen/Documents/Bibtex/Bibtex_database}

\end{document}